\documentclass[twocolumn]{aastex61}

\shorttitle{LORENTZ FACTORS OF BHXRB COMPACT JETS}
\shortauthors{Saikia et al. 2019}
\usepackage{rotating}
\usepackage{bm}
\usepackage{amsmath}
\begin{document}

\title{LORENTZ FACTORS OF COMPACT JETS IN BLACK HOLE X-RAY BINARIES}

%% AUTHOR/INSTITUTIONS FOR AASTEX6.1:
\author{Payaswini Saikia}
\affiliation{Center for Astro, Particle and Planetary Physics, New York University Abu Dhabi, PO Box 129188, Abu Dhabi, UAE}

\author{David M. Russell}
\affiliation{Center for Astro, Particle and Planetary Physics, New York University Abu Dhabi, PO Box 129188, Abu Dhabi, UAE}

\author{D. M. Bramich}
\affiliation{Center for Astro, Particle and Planetary Physics, New York University Abu Dhabi, PO Box 129188, Abu Dhabi, UAE}
\affiliation{Center for Space Science, NYUAD Institute, New York University Abu Dhabi, PO Box 129188, Saadiyat Island, Abu Dhabi, UAE}
\affiliation{Division of Engineering, New York University Abu Dhabi, PO Box 129188, Saadiyat Island, Abu Dhabi, UAE}

\author{James C. A. Miller-Jones}
\affiliation{International Centre for Radio Astronomy Research - Curtin University, GPO Box U1987, Perth, WA 6845, Australia}

\author{Maria Cristina Baglio}
\affiliation{Center for Astro, Particle and Planetary Physics, New York University Abu Dhabi, PO Box 129188, Abu Dhabi, UAE}
\affiliation{INAF, Osservatorio Astronomico di Brera, Via E. Bianchi 46, I-23807 Merate (LC), Italy}

\author{Nathalie Degenaar}
\affiliation{Anton Pannekoek Institute for Astronomy, University of Amsterdam, Postbus 94249, 1090 GE Amsterdam, The Netherlands}
%\affiliation{More affiliations}

%% AUTHOR/INSTITUTIONS FOR EMULATE APJ:
% \author{Patricio~E.~Cubillos\altaffilmark{1,2},
% Joseph~Harrington\altaffilmark{1},
% and
% Third~Author\altaffilmark{1}
% }
% \affil{\sp{1} Planetary Sciences Group, Department of
%               Physics, University of Central Florida, Orlando, FL 32816-2385\\
%        \sp{2} Space Research Institute, Austrian Academy of Sciences,
%               Schmiedlstrasse 6, A-8042, Graz, Austria}

\email{ps164@nyu.edu}

% %% Extra info for aastex:
% \received{Yesterday}
% \revised{Today}
% \accepted{Tonight}
% \published{Tomorrow}
% \submitjournal{AASJournal}

\begin{abstract}
Compact, continuously launched jets in black hole X-ray binaries (BHXBs) produce radio to optical-infrared synchrotron emission. In most BHXBs, an infrared (IR) excess (above the disc component) is observed when the jet is present in the hard spectral state. We investigate why some BHXBs have prominent IR excesses and some do not, quantified by the amplitude of the IR quenching or recovery over the transition from/to the hard state. We find that the amplitude of the IR excess can be explained by inclination dependent beaming of the jet synchrotron emission, and the projected area of the accretion disc. Furthermore, we see no correlation between the expected and the observed IR excess for Lorentz factor 1, which is strongly supportive of relativistic beaming of the IR emission, confirming that the IR excess is produced by synchrotron emission in a relativistic outflow. Using the amplitude of the jet fade and recovery over state transitions and the known orbital parameters, we constrain for the first time the bulk Lorentz factor range of compact jets in several BHXBs (with all the well-constrained Lorentz factors lying in the range of $\Gamma$ = 1.3 - 3.5). Under the assumption that the Lorentz factor distribution of BHXB jets is a power-law, we find that N($\Gamma$) $\propto \Gamma^{ -1.88^{+0.27}_{-0.34}}$. We also find that the very high amplitude IR fade/recovery seen repeatedly in the BHXB GX 339-4 favors a low inclination angle ($\lesssim 15^\circ$) of the jet. 

\end{abstract}

% http://journals.aas.org/authors/keywords2013.html

\section{INTRODUCTION}
\label{introduction}

Accreting compact objects on a wide range of physical scales often produce powerful relativistic, collimated outflows known as jets. Astrophysical jets exist in both supermassive and stellar-mass black holes, and seem to share many common features. Relativistic jets powered by stellar-mass black hole X-ray binaries (BHXBs) typically manifest either as steady, continuous, compact outflows during the hard X-ray state or as discrete, ballistic, super-luminous ejections during state transitions \citep[eg.][]{fe04}. The compact jets observed during the hard state of a BHXB are very similar to the relativistic compact jets produced in supermassive black hole systems i.e. Active Galactic Nuclei (AGN). Both of them share the signature flat-to-inverted ($\alpha$ $>$ 0-0.5, where F$_{\nu} \propto \nu^{\alpha}$) radio spectrum, extending up-to infrared and sometimes even optical wavelengths. This emission is associated with partially self-absorbed synchrotron radiation originating at a steady, unresolved, compact jet \citep{bk79}. A tight correlation has been found between near-simultaneous observations of the jet radio luminosity and the X-ray luminosity in BHXBs in the hard state \citep[eg.][]{g03,c03,g14}. The same relation also extends to AGN through the X-ray and optical Fundamental Plane of black hole activity \citep[eg.][]{m03,f04,s18}, suggesting scale invariance of compact, relativistic jets.\\

The observed luminosity of the jet is affected by relativistic beaming, which depends on the bulk Lorentz factor and the viewing angle, or the inclination of the source. A clear understanding of these intrinsic properties of a jet is crucial for constraining the physics of the jet launching region, collimation and acceleration of jets, and the intrinsic physical properties of the black holes. There are many studies that have estimated the bulk Lorentz factor for jets in AGN and found a parent distribution in the form of a power law \citep[eg.][]{pu92,lm97,k04,j05,s16}. However, the bulk Lorentz factor in BHXB jets is notoriously difficult to measure, with to date only weak constraints for a few BHXBs. While \cite{g03} used the scatter in the radio/X-ray relation of BHXB jets to constrain the Lorentz factors of compact jets to $\Gamma <$2, \cite{hm04} showed that that this correlation does not exclude high values of Lorentz factors. Meanwhile for transient jets launched during state transitions, \cite{f03} argued that one can only estimate a lower limit on the Lorentz factors by using two-sided jet proper motions. \cite{miller06} used opening angles of transient jets to estimate the bulk Lorentz factor of these sources to have a mean $\Gamma$ $>$10, assuming that the observed opening angles are due to the transverse Doppler effect.\\

In this paper, we adopt a simple model to constrain the Lorentz factor of the compact jets in several BHXBs, using infrared (IR) observations. For some BHXBs, the accretion disc tends to dominate the IR emission throughout the outbursts. However, for many sources (eg. XTE J1550-564 \citep{jain01}, 4U 1543-47 \citep{bb04}, H1743-322 \citep{chaty15}, XTE J1650-500 \citep{curran} and GX 339-4 \citep{cf02,ho05}), there is an IR excess owing to synchrotron emission produced in the jets. 

It has been found that the IR excess above the disc component is prominent in the hard state, with a correlation between IR and X-ray luminosities similar to the radio/X-ray correlation \citep[eg.][]{ho05,r06,cor,vin}. The IR excess is also absent in the soft state, following the behaviour of the radio jet \citep[eg.][]{mil,kal}. The IR emission in the soft state is dominated by the accretion disc \citep[eg.][]{ho05}. The IR excess appears to fade and recover close to the transition away from the hard state \citep{cor,cad,cristina}, whereas the radio emission can persist over the transitions, existing in the intermediate states \citep[eg.][]{fe04,mi12}. This was found to be due to an evolving jet spectrum \citep{c13,van,rus14b}, whereby the spectral break (between the partially self-absorbed, optically thick synchrotron spectrum and the optically thin spectrum) is correlated with the X-ray hardness over the transition, and this correlation even exists in low-luminosity AGN \citep{koko}. Indeed, the IR excess in the hard state has an optically thin synchrotron spectrum \citep[eg.][]{gandhi,rah12,dave13,cristina} and, although alternative, inflow models have been proposed for the origin of the IR excess \citep{pou}, it has been shown that the general properties are more in line with a jet origin \citep[eg.][]{kal}.

To date, there is no explanation as to why some sources appear to have prominent IR excesses \citep[eg. GX 339-4,][]{cf02,ho05}, while others do not \citep[eg.XTE J1720-318,][]{cb06}. It has been shown, however, that sources that are IR-faint on the IR/X-ray correlation are also radio-faint on the radio/X-ray correlation \citep{curran,dave13,chaty15}. Since the jets are relativistic, we expect the IR emission to be Doppler boosted, i.e. beamed towards or away from the observer. Therefore, sources with jets pointing towards our line of sight may have more prominent IR excesses than those directed away. If the jet axis is perpendicular to the orbital plane (see discussion below), then one can test for a correlation between the inclination angle and the jet emission using measured inclination angles. This was recently tested by \cite{motta} using radio data -- they found that sources with low inclination angles appear, in most cases, to be radio-loud on the radio/X-ray correlation, whereas higher inclination sources tended to be radio-quiet. The IR excess will also be affected by the emission from the accretion disc. Larger discs, and face-on discs (low inclination), will tend to decrease the jet/disc emission ratio. For neutron star X-ray binaries, which also have IR excesses in some cases, it was suggested that systems with small discs have more prominent IR excesses \citep{r07}.\\

Here, we explore whether relativistic beaming, and relative disc emission, can explain the observed IR excess in BHXBs as measured from the amplitude of the IR fade and recovery over state transitions. We describe our sample of black hole X-ray binaries in Section 2, and gather from the literature the physical properties of the sources as well as infrared data during outbursts. In Section 3, we discuss the model and the methods used to estimate the expected IR flux excess in these sources from their observed parameters like the masses involved, orbital period, inclination angle, etc. In Section 4 we present the Bayesian framework used to constrain the jet Lorentz factors for our sample of BHXBs. Finally, in Section 5 we discuss the results of the paper, caveats involved in our method, and compare our findings with previous literature.

\section{SAMPLE}
\label{sec:sample}

About 77 BHXBs are known in the Galaxy \citep[though only 21 of these are dynamically confirmed as hosting black holes;][]{teta}. We have compiled a list of 14 BHXBs for which reliable IR excess information is available in the literature. In this section, we describe the methods used to estimate the IR excess and inclination angles. We also present the sources and caveats of the infrared observations, and discuss the black hole properties used in this study.\\

\textit{Methods of estimating IR excess :} We use three methods to measure the IR excess over disc emission. When well-sampled near-infrared (NIR) K-band (or H-band) light curves are available, we estimate the amplitude of the IR excess by simply measuring the magnitude change observed over the transition. This transition usually occurs fairly quickly, with a clear IR fade or rise lasting days to a week or so, whereas the outer disc component in the IR is largely unchanged over this short period of time. Hence this method provides a good measure of the IR jet quenching and recovery \citep[eg.][]{bb04,r10,dave1752}. For sources that do not make state transitions (e.g. sources that remain in the hard state), we estimate the amplitude of the IR excess above the disc component from published optical--IR spectral energy distributions (SEDs), where the disc and jet components are both measured. Finally, in one case H1743--322, we infer the IR excess amplitude from the change in the IR $J$--$K_{\rm s}$ colour between hard state and soft state observations (see section 2.5 for details).

In order to estimate the contribution of the companion star to the total IR flux, we compare the companion's quiescence IR flux to the observed flux of the system in all cases. We find that in only 1 source, GRO J1655-40, the companion flux exceeds 10\% of the observed flux, and hence needs to be taken care of (see Section 2.7 for details). For all the other sources in our sample, we find that the the IR contribution of the companion is much less than 10\% of the total IR flux of the system.\\

We chose to use the IR flux change as a ratio (flux with jet on / flux with jet off) rather than an absolute flux value. We do this to be able to compare between sources, because the flux values depend on the distance, whereas the ratio does not. Distances are highly uncertain in many cases. This is also one of the reasons why we have chosen not to use radio data in this work. The disc does not contribute to the radio emission, and the non-jet radio emission in the soft state is approximately zero. We therefore cannot use a jet on/off flux ratio if we were to use radio data, and would be forced to use the absolute flux values, requiring the distance which introduces errors. In addition, radio data over the state transition are complicated by bright transient ejections, and monitoring of the compact, flat spectrum jet over the transitions is poorly sampled in many cases. In a future study, radio data could be used for sources with well known distances and well-sampled light curves, to infer the bulk Lorentz factors of jets at (radio-emitting) large distances from the black hole. By using IR data here, we are probing the Lorentz factor of jets close to their launching region \citep[at distances on the order of $\sim 0.1$ light-seconds from the black hole;][]{gandhi2017}.\\

\begin{table*}
\centering
  \begin{tabular}{ | l | r | r | r | r | r | r | r |}
\hline
Name & Inclination   &  $M_{\rm BH}$  & $M_{\rm CS}$  &  Orbital Period & Distance& IR excess & IR \\ 
 & ($^\circ $)  &   ($M_{\odot}$) & ($M_{\odot}$) &   (hrs) &  (kpc) &  (mag) & IR \\ 
\hline
XTE J1118+480	& 68 - 82	& 7.3$\pm$0.7 &	0.18$\pm$0.07	& 4.078414$\pm$5$\times$10$^{-6}$	&	1.7$\pm$0.1	&	0.24-3.10* & K\\
Swift J1357.2-0933	& 80 - 90	& $>$ 9.3		& 0.4$\pm$0.2 & 	2.8$\pm$0.3		& 1.5	 - 6.3		& 1.00$\pm$0.13  & K\\
\textit{MAXI J1535-571} &  --	& 7.7 - 10.0 &	--	& --	&	--	&	2.10$\pm$0.16  & K\\
4U 1543-47 &  20.7$\pm$1.5 &  9.4$\pm$1.0 &  2.45$\pm$0.15 &  26.79377$\pm$7$\times$10$^{-5}$	 & 9.1$\pm$1.1	& 1.87$\pm$0.03  & K\\
XTE J1550-564 & 57.7 - 77.1 & 9.1$\pm$0.6   &  0.30$\pm$0.07 & 37.008799 $\pm$ 5.8$\times$10$^{-5}$	&	4.38$^{+0.58}_{-0.41}$ &0.97$\pm$0.07  & H\\
XTE J1650-500 & 75.2$\pm$5.9 & 4.7$\pm$2.2  &  $<$ 2.36 & 7.69$\pm$0.02 & 2.6$\pm$0.7	& 0.53$\pm$0.18  & K\\
GRO J1655-40	& 70.2$\pm$1.9	& 5.4	$\pm$0.3		& 1.45$\pm$0.35	& 62.9258$\pm$4.8$\times$10$^{-3}$	&	3.2$\pm$0.5	&	0.00 - 0.24  & K\\
\textit{GX 339-4	} &	0 - 78	& 2.3 - 9.5		& 0.41-1.71 & 	42.14$\pm$0.01	&	6 - 15	&	1.50 - 3.20$^{+}$ & H\\
\textit{H 1743-322}	& 75$\pm$3	& -- &	--	& --	&	8.5$\pm$0.8 &	0.97$\pm$0.12  & K\\
\textit{XTE J1752-223} 	& $<$ 49	& 9.6$\pm$0.9 &	--	& $<$ 22	&	3.5$\pm$0.4	&	0.35$\pm$0.18  & H \\
Swift J1753.5-0127	& 40 - 80	& $>$7.4	& 0.17 - 0.25	& 2.85-3.24  &	1 - 10		& 0.0 - 0.6   & K\\
MAXI J1836-194 & 4 - 15  &  $>$ 2.0 &  $<$ 0.65 & $<$ 4.9 & 4 - 10	&	2.38$\pm$0.43 & K \\
XTE J1859+226	 & 60$\pm$3	& 10.8$\pm$4.7		& $<$ 5.41	& 6.58$\pm$0.05	&	8$\pm$3	&	0.68$\pm$0.03   & K\\
\textit{Swift J1910.2-0546}	& --	& $>$ 2.9 &	--	& 2.2 - 4.0	&	$>$ 1.70	&	0.41$\pm$0.28  & K\\
\hline
\end{tabular}
\caption{Physical properties and orbital parameters obtained from the literature, for all the BHXBs which have previous measurements/estimates of infrared excess observed during state transitions. The columns include the name of the source, inclination angle, mass of the black hole, mass of the companion star, orbital period, distance, IR excess and the IR band in which the excess has been measured, respectively. For each of the sources, the values obtained from the literature and their references are discussed in the text. We do not include the sources in italics in our Bayesian analysis because their parameters are not well-measured, however we include them in the table for completeness and potential future studies. Values in the table have been rounded to a uniform level of precision. (`--' : no measurement is available, `*' : model-dependent values of IR excess explained in the text, `+'  : many measurements available in the literature are explained properly in Table 2).}
\end{table*}

\textit{Methods of deciding inclination constraints :} At present, there are mainly three ways to estimate the inclination angle of a BHXB  - (i) using the orbital inclination of the companion star to our line-of-sight as typically measured from ellipsoidal modulation of the companion light curve in quiescence, (ii) using the inclination of the inner disc close to the central black hole as typically measured from X-ray spectral fitting, and (iii) using the inclination of the jets as measured from the relative fluxes and sizes of the two jets estimated from radio observations. We have chosen to exclude inclinations constrained via X-ray spectral fitting because they are highly model dependent and can give wildly differing values depending on the assumptions and models used \citep[see for eg.][]{hiemstra}. Moreover, there is increasing evidence that at least in some sources the inner disk is precessing \citep[see for eg.][]{ogli,liska18,motta18,james19} and perhaps that causes (some of) the discrepancy between inclinations measured through X-ray reflection models. On the other hand, the discrepancy in inclinations estimated from optical measurements is usually only a few degrees \citep{cryno}. For uniform selection and reliable measurement of inclination, we only consider inclinations derived either from optical measurements, or through radio observations when optical measurements are unavailable.

\subsection{XTE J1118+480}

XTE J1118+480 was discovered by the All Sky Monitor (ASM) on the RXTE (Rossi X-ray Timing Explorer) satellite in 2000 by \cite{rem}. The inclination measurements reported by different studies all lie in the range $68^\circ$-$82^\circ $ \citep[eg.][]{wag,mcc,zur,cryno}. \cite{teta} established a comprehensive database of BHXBs named the Whole-sky Alberta Time-resolved Comprehensive black-Hole Database Of the Galaxy where they calculated a central black hole mass of 7.30 $\pm$ 0.73 $M_{\odot}$, and tabulated a distance of 1.7 $\pm$ 0.1 kpc \citep{gelino}, a mass ratio between the companion star and the central black hole of 0.024$\pm$0.009 \citep{calvelo} and an orbital period of 4.078414$\pm$0.000005 h \citep{torres}.

\cite{hihi} and \cite{hihi2006} observed this source in the IR band, but because it remained in the hard state throughout both its outbursts, it is difficult to separate out the disc and jet components. \cite{mcc} modeled the optical/UV with a multi-temperature disc and found that the NIR band fluxes needed to be multiplied by 0.8 to lie on the extrapolation of the optical disc spectrum. So the jet component "factor flux drop" can be estimated as 1/0.8 = 1.25, which is a change of $\sim$0.24 mag. However, \cite{chaty2003} modelled the broadband SED of the source and found the K-band flux to be 1.24 orders of magnitude brighter than the disc component. Therefore, the jet component "factor flux drop" can be estimated as $10^{1.24}$ $\approx$ 17.33, which is a change of $\sim$3.10 mag. Clearly the jet contribution is model dependent, and hence we use a range of IR excess to incorporate both the values in our final analysis to estimate the Lorentz factor.

\subsection{Swift J1357.2-0933}

Swift J1357.2-0933 was detected in 2011 by the Swift Burst Alert Telescope \citep{cream}. \cite{corral} performed time-resolved optical spectroscopy of broad, double-peaked H$\alpha$ emission in this BHXB, and estimated an orbital period of 2.8$\pm$0.3 hrs. \cite{mata15} studied this source during quiescence, and estimated a massive black hole with $M_{BH}$ $>$ 9.3 $M_{\odot}$, a companion star with mass $\sim$ 0.4 $M_{\odot}$ (we conservatively assume an error of 50\%, i.e. $M_{\rm CS}$ = 0.4$\pm$0.2 $M_{\odot}$ ), and a very high orbital inclination ($i$ $>$ 80$^\circ$). The distance to the source ranges from $\sim$1.5 - 6.3 kpc \citep{rau,shah}.

\cite{shah} performed high time resolution ULTRACAM optical and NOTCam infrared observations of Swift J1357.2-0933 during the 2011 outburst. They showed that during the 2011 outburst the K-band flux was $\sim$ 0.4 $\pm$ 0.05 dex brighter than the disc model that was well fit to the SED. This is a factor of 2.51 $\pm$ 1.12 flux change, or a 1.00$\pm$0.13 mag change.\\

\subsection{MAXI J1535- 571}

MAXI J1535-571 was discovered by MAXI in 2017 \citep{negoro}. \cite{shang} performed X-ray spectral analysis of the source and estimated the mass of the black hole to be in the range of 7.7 - 10.0 $M_{\odot}$. The inclination of the source is not properly constrained, with reports of estimated measurements from model-dependent X-ray spectral fitting ranging from 27$^\circ$ to 67$^\circ$ \citep[eg.][]{gen,xuxu,sk18}. As different values of inclinations for this source in the literature are all from X-ray measurements and they are in disagreement with each other by more than 30 degrees, we consider the inclination of this source to be very uncertain and we do not use it for our study.

The NIR ($JHK$) and optical ($yzir$) light curves of MAXI J1535-571 during its 2017/2018 outburst are reported in \citep{cristina}. A clear fading is observed over the hard to soft transition, with the drop in flux being more significant towards the lowest frequencies. The magnitude drop measured from the light curve is 2.10$\pm$ 0.16 mag in the K-band. 

\subsection{4U 1543-47}

4U 1543-47 is a recurrent X-ray transient first discovered in 1971 \citep{72}. \cite{oro98} performed spectroscopic observations of the source and found the distance to the source to be $\sim$ 9.1$\pm$1.1 kpc, assuming the secondary to be on the main sequence. The inclination of the source was found to be 20.7$^\circ$$\pm$1.5$^\circ$ \citep{oro03}. \cite{r06} compiled from the literature a central black hole mass of 9.4$\pm$1.0 $M_{\odot}$ and a companion mass of 2.45$\pm$0.15 $M_{\odot}$. The orbital period of the source is 26.79377$\pm$0.00007 hrs \citep{oro03}.

\cite{bb04} studied re-brightening of the K-band light curve and reported a K-band magnitude change from 14.06$\pm$0.03 mag on MJD 52473.221100 to a value of 12.19$\pm$0.01 mag on MJD 52487.961500. This implies a change of 1.87$\pm$0.03 mag during state transition.

\subsection{XTE J1550-564}

XTE J1550-564 was discovered by RXTE/ASM \citep{smith99}. \cite{oro11} used moderate-resolution optical spectroscopy and near-infrared photometry of the source to find an orbital period of 1.5420333 $\pm$ 0.0000024 days (37.008799 $\pm$ 5.8$\times$10$^{-5}$ h). Using the light curves obtained, they estimated an inclination range of 57.7$^\circ$ - 77.1$^\circ$, a black hole mass of 9.10$\pm$0.61 $M_{\odot}$, a secondary star mass of 0.30$\pm$0.07 $M_{\odot}$, and a distance of 4.38$^{+0.58}_{-0.41}$ kpc.

H-band light curves of XTE J1550-564 were reported by \cite{jain01} and also analysed in \cite{r07,r10,r11}. These observations showed that during the hard to soft transition, the H-band faded from 13.3815 $\pm$ 0.0225 mag to 14.3365 $\pm$ 0.0645 mag (i.e. a change of 0.955 $\pm$ 0.068 mag). The soft to hard transition showed a H-band rise from 14.8770 $\pm$ 0.0750 mag to 13.8905 $\pm$ 0.0135 mag (i.e. a change of 0.987 $\pm$ 0.076 mag). For this analysis, we use the combined (averaged) value of $\sim$0.97 $\pm$ 0.07 mag change.

\subsection{XTE 1650-500}

XTE J1650-500 was discovered in September 2001 by the RXTE/ASM \citep{rem01}. \cite{ororo} used optical observations to derive an orbital period of 0.3205$\pm$0.0007 days ($\sim$ 7.69$\pm$0.02 h) and estimate the upper-limit mass of the central black hole to be 7.3 $M_{\odot}$. \cite{cryno} estimated an inclination of $i$ = 75.2$^\circ \pm$5.9. \cite{teta} calculated a central black hole mass of 4.72 $\pm$ 2.16 $M_{\odot}$, estimated a mass ratio between the companion star and the central black hole in the range 0.0-0.5, and tabulated a distance of 2.6 $\pm$ 0.7 kpc \citep{homan}.

\cite{curran} showed that there is a clear IR drop over the hard to soft transition. There is no light curve of the drop itself, but it fades from K$_s$ = 13.29$\pm$0.13 on MJD 52161.03691 to K$_s$ = 13.82$\pm$0.12 on MJD 52177.00145, showing an IR drop of 0.53$\pm$0.18 mag over the transition. It is important to note that there is only one data point in the hard state. However the hard state point is just before (within a day) the hard to intermediate state transition when the IR starts to fade in all sources with a well sampled light curve. So the drop of 0.53$\pm$0.18 mag is likely to be accurate within errors.

\subsection{GRO J1655-40}

GRO J1655-40 was discovered in 1994 by BATSE on board CGRO \citep{harmon95}. \cite{hr95} estimated the distance to the system to be 3.2$\pm$0.5 kpc. \cite{greene} performed BVIJK photometry of the source during full quiescence, and found an orbital period of 2.62191$\pm$0.00020 days (or 62.9258$\pm$0.0048 h), inclination angle of 70.2$^\circ$ $\pm$ 1.9$^\circ$ and a black hole mass of 6.3$\pm$ 0.5 $M_{\odot}$. Later, \cite{beer} estimated the mass of the black hole to be 5.4$\pm$ 0.3 $M_{\odot}$, and the mass of the companion star to be 1.45$\pm$ 0.35 $M_{\odot}$.

From \cite{kalemci}, we estimate the hard to hard-intermediate state transition to have happened around MJD 53435 during its 2005 outburst. \cite{shid} and \cite{kalemci} showed no visible change in K-band at this time, with flux changing by a factor of less than 1.25 (0.24 mag). \\

One peculiar thing about GRO J1655-40 is that unlike the other sources in our sample, the contribution of companion star to the IR flux of the system is more than 10\%. From \cite{kalemci}, the average magnitude of the system in the 5 days preceding the transition is 12.222$\pm$0.029 mag (F$_1$ = 8.616 mJy). The average magnitude of the system 5 days after the transition (including the transition day) is 12.228$\pm$0.059 mag (F$_2$ = 8.569 mJy). The K-band magnitude of the system during quiescence is 13.3 mag (F$_{\rm Q}$ = 3.193 mJy). So the companion star produces $\sim 37\%$ of the flux in this system at the time of the transition. Removing the IR flux contribution of the companion star (which is constant over the transition), we find that the the flux ratio becomes (F$_1$-F$_{\rm Q}$)/(F$_2$-F$_{\rm Q}$) = 1.009. This corresponds to a magnitude change of $\sim$0.01 mag, which is much smaller than the IR flux change limits (0--0.24 mag) we have used in the above analysis. We therefore adopt the more conservative range, 0--0.24 mag as the minimum and maximum flux drop over the transition.

\subsection{GX 339-4}

\begin{table}
\centering
  \begin{tabular}{|c|c|c|}
\hline
Year & MJD during transition & IR Mag change\\ 

\hline

2002 (rise) & 2390.7649 - 2405.8173	&	3.11$\pm$0.028 \\
2003 (fade) & 2738.82521 - 2759.83738	& 1.51$\pm$0.036  \\
2004 (rise) & 3217.6421 - 3231.62454	& 1.56$\pm$0.028  \\
2005 (fade) & 3476.84335 - 3491.85172 & 1.65$\pm$0.036 	 \\
2007 (rise) & 4134.86885 - 4147.84173	& 3.20$\pm$0.028 	 \\
2007 (fade) & 4239.80197 - 4254.79423	& 1.50$\pm$0.036 	 \\
2010	 (rise) & 5293.8583 -5302.8286 &  2.98$\pm$0.036 	\\
2011 (fade) & 5605.88975 - 5617.77736	& 1.64$\pm$0.036 	 \\
\hline
\end{tabular}
\caption{IR excess measured in different state transitions of GX 339-4 during 2002-2011. IR rise is measured during the transition towards the hard state, while a fade is observed during the transition away from the hard state.}
\end{table}

\begin{figure}
\center
\includegraphics[height=0.33\textwidth]{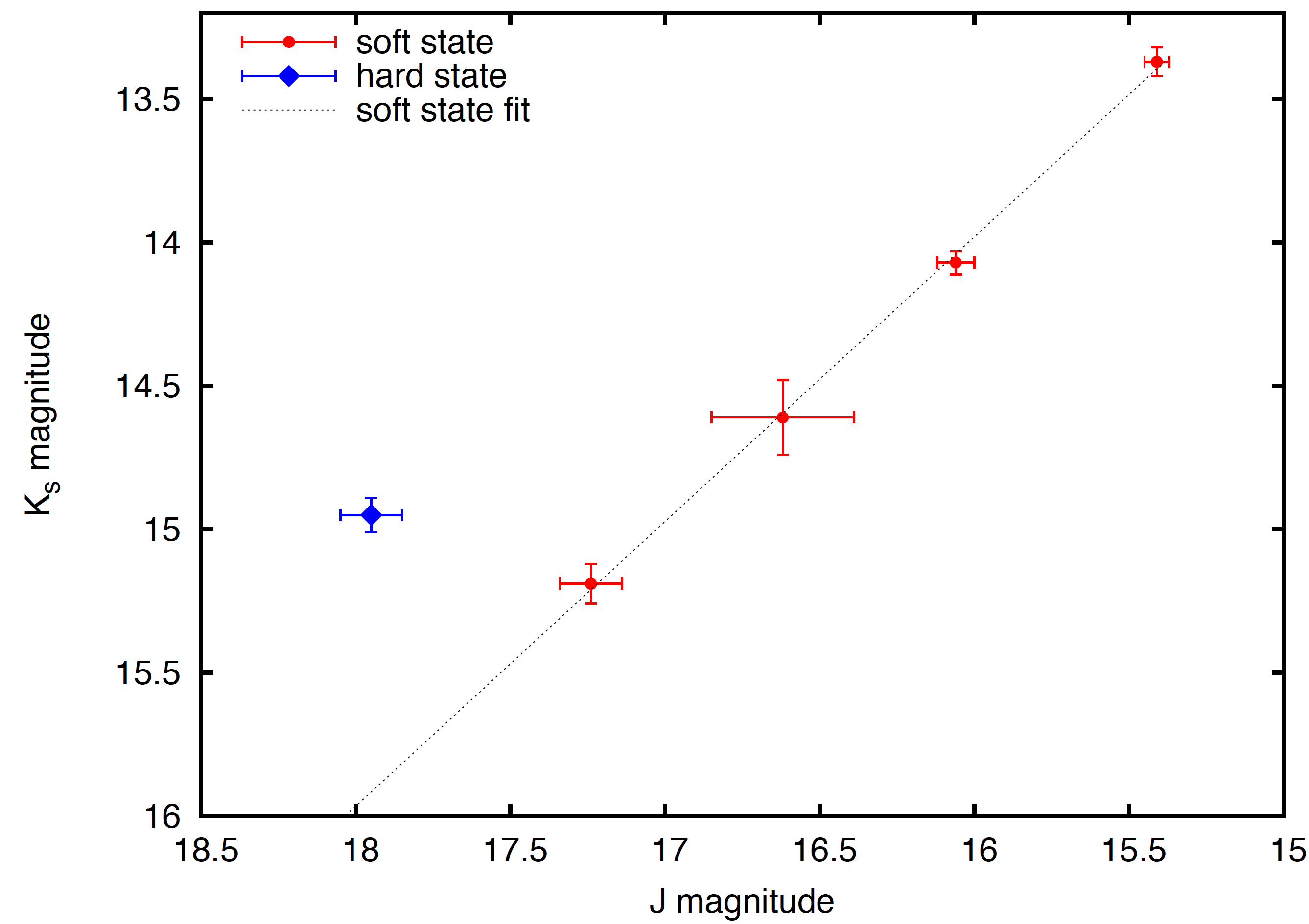}
\caption{IR observation of H1743 during soft (red) and hard (blue) states from \citep{chaty15} (see text for explanation).}
\end{figure}

GX 339-4 was discovered in 1972 by the MIT X-ray detector on board the Orbiting Solar Observatory (OSO) 7 satellite. It shows relatively frequent outburst cycles of various strengths (typically once every 2-3 years). The orbital period of the system is estimated to be 1.7557$\pm$0.0004 days ($\sim$ 42.14$\pm$0.01 hrs) by measuring Doppler shifts of fluorescent lines \citep{hygx}. The distance to the source is expected to be in the range of 6-15 kpc \citep{hy04}. However, the mass of the compact object and companion star, as well as the inclination of the system are very weakly constrained. \cite{h17} analysed the radial velocity curve and projected rotational velocity of the donor to set limits of 2.3 $M_{\odot}$ $\le$ $M_{BH}$ $\le$ 9.5 $M_{\odot}$ to the accretor mass, and a mass ratio $q$ = 0.18 $\pm$ 0.05. The inclination angle of GX 339-4 is not well-constrained and values in the literature range from 13$^\circ$$\pm$3$^\circ$ \citep[][from X-ray]{m2004} and 0$^\circ$-30$^\circ$ \citep[][from optical]{wu2001} to 50$^\circ$$\pm$10$^\circ$ \citep[][from X-ray]{basak} and 57.5$^\circ$$\pm$20.5$^\circ$ \citep[][from optical]{h17}. Owing to the high uncertainty in the inclination angle, we do not include this source in our final analysis. \\

GX 339-4 has a good coverage of IR excess measured during different transitions over the years (see Table 2) due to intensive monitoring by SMARTS \citep{cor,bux,din}. All the IR excess magnitudes reported for GX 339-4 are roughly in two different ranges, 1.50-1.65 mag and 2.98-3.20 mag. We discuss the possible reasons behind these two observed IR excess values, any relation of the differences with the type of transition, and their correlation with the X-ray luminosity in the discussion section. 

\subsection{H1743-322}

H1743-322 was discovered during an outburst in 1997 by the Ariel V satellite \citep{kholt} and HEAO-1 satellite \citep{dox}. \cite{stei} used a kinematic model of the jets to estimate the distance to this source as 8.5$\pm$0.8 kpc and the inclination angle of the jets as 75$^\circ$ $\pm$3$^\circ$. In the literature, there is no inclination measurement using optical data. Presently, the black hole mass, companion star mass and the orbital period of this system are also not clearly known.

A direct measurement of the IR excess from the light curve is not available for this source, because its outbursts were only partly sampled in the IR. However, \citet{chaty15} published IR magnitudes of the source from several outbursts. We find that there is a strong correlation between the $J$ and $K_{\rm S}$-band magnitudes in the soft state (see Fig. 1), which is very likely due to disc emission. In the hard state (when we expect a jet contribution), the data deviate from the correlation; the $K_{\rm S}$-band being brighter than expected from the relation, by $0.97 \pm 0.12$ mag. This is therefore a measure of the $K_{\rm S}$-band excess due to the jet emission, above the disc component, so we take this as the magnitude change over the state transition.

\subsection{XTE J1752-223}

XTE J1752-223 was discovered in 2009 by RXTE \citep{mark}. \cite{shap10} used correlations between spectral and variability properties with GRO J1655-40 and XTE J1550-564, and estimated a distance of 3.5$\pm$ 0.4 kpc and a BH mass of $9.6\pm 0.9\;{M}_{\odot }$. \cite{dave1752} performed optical monitoring of this source during its 2009-10 outburst and decay to quiescence, and estimated a likely orbital period of $<$ 22 hrs. \cite{milj} measured the inclination angle of the source to be $<$ 49$^\circ$.

\cite{chun1752} performed simultaneous X-ray and optical/near-infrared observations of the source during its outburst decay in 2010. They showed that over the transition from the soft state towards the hard state, the H-band emission increased by a small amount : 0.35 $\pm$ 0.18 mag.

\subsection{Swift J1753.5-0127} 

Swift J1753.5-0127 was discovered in outburst by Swift/BAT in 2005 \citep{palmer}. It had a $\sim$12 year long outburst, with a mini-outburst towards the end \citep{plot,shaw2019}. \cite{distz} estimated the photometric orbital period of the source to be 3.2443$\pm$0.0010 h. Later, \cite{neu14} used radial velocity measurements to estimate an orbital period of 2.85$\pm$0.01 hrs, which is one of the shortest orbital periods of any known BHXB. For our calculation, we use an orbital period range of 2.85-3.24 hrs to incorporate both values. Using high-resolution, time-resolved optical spectroscopy, \cite{shaw16} estimated the mass of the compact object to be $>$7.4$\pm$1.2 $M_{\odot}$. Although the mass of the companion star is not known, \cite{neu14} estimated it to be in the range 0.17-0.25 $M_{\odot}$ using empirical and theoretical mass-period relations for a 2.85 hrs orbital period binary. The inclination of the source is not precisely known; while \cite{neu14} suggests a lower limit of 40$^\circ$, \cite{shaw2019} predicts an upper limit of 80$^\circ$. The distance to this source is also not well constrained and different studies report a large range of possible distances of 1-10 kpc \citep{distc,distz,distf}. 

\cite{tom} carried out a multi-wavelength campaign of Swift J1753.5-0127 in the hard state during 2014 April. The K-band flux was measured as $\sim$ 0.4-0.7 mJy, while the jet emission was $\sim$0.08 mJy (estimated from Fig. 8(b) of \cite{tom}). Using these estimates, we can assume a factor change in the range of 1.13-1.25 when the jet switches off, although we cannot rule out the possibility that the jet could be a lot fainter. In \cite{rah15}, the K-band flux is calculated to be $\sim$0.7 mJy and the disc emission is measured as $\sim$0.3 mJy. So if the jet would switch off, a factor change of 1.75 would be expected. Based on these two studies, we can safely assume a factor flux change in the range of 1-1.75 (equivalently 0-0.6 mag).

\subsection{MAXI J1836-194}

MAXI J1836-194 was discovered in the early stages of its outburst in 2011 \citep{nego11}. \cite{rus14} studied the Very Large Telescope optical spectra of the source, and estimated an inclination angle between 4$^\circ$ - 15$^\circ$, assuming distances between 4 and 10 kpc. The donor is expected to be a main-sequence star with a mass $<$0.65 $M_{\odot}$, with an orbital period of $<$4.9 h. The mass of the compact object is not well-constrained. However we can estimate a lower limit of 2 $M_{\odot}$, given the lower limit on the distance of 4 kpc.

Data from \cite{ab,rus14} showed that there is an IR magnitude change during state transition from 13.86$\pm$0.37 mag on MJD 55820.12 to 11.48$\pm$0.21 mag on MJD 55845.0511. This shows a rise of IR emission by 2.38$\pm$0.43 mag in $\sim$24.9 days.\\

\subsection{XTE J1859+226}

XTE J1859+226 was first detected by RXTE in 1999 \citep{wood99}. \cite{11cs} used both optical photometry and spectroscopy to find an orbital period of 6.58 $\pm0.05$ h, and an inclination angle of $60^{\circ} \pm 3^{\circ}$. \cite{teta} has tabulated a central black hole mass of 10.83 $\pm$ 4.67 $M_{\odot}$ and a companion star upper limit of $<$ 5.41 $M_{\odot}$. The distance to this source is still not certain, with different studies inferring statistically different distances, for example 11 kpc \citep{zu02}, 4.6-8.0 kpc \cite{hyhyhy} and 8$\pm$3 kpc \cite{hynesk}. We adopt the latter for our calculation as it encompasses all the values reported.

\cite{hy2002} observed the source in J, H and K bands during 1999-2000 and showed that there is a small IR excess in the lowest frequencies, at the beginning of the state transition from the hard state to the soft state. As shown in \cite{bro}, the initial hard state is until MJD $\sim$51464, after which the source made the transition. The first NIR data in \cite{hy2002} is on MJD 51465, at the start of the transition. The K-band NIR excess can be seen in the first SED on MJD 51465, where the J-band data point is close to the disc spectrum \cite[see fig 4,][]{hy2002}. Comparing that with the SED obtained on MJD 51469, when the K-band lies on the disc extrapolation from optical/UV, we can say that the jet has faded while the disc remained. The K-band drop during this period is calculated as 0.68$\pm$0.03 mag.

\subsection{Swift J1910.2-0546}

Swift J1910.2-0546 was simultaneously discovered by Swift/BAT \citep{krimm} and MAXI \citep{usui} in 2012. \cite{1} and \cite{2} examined periodic variations in the optical light curve to estimate the orbital period to be $\sim$2.2 hrs and $\sim$4 hrs, respectively. \cite{naka} did long-term monitoring of the source and put a lower limit on the mass of the compact object of $>$2.9$M_{\odot}$ and on the distance of  $>$1.70 kpc. The inclination angle is estimated to be between 4 - 22 degrees from X-ray spectral fitting \citep{rrr}. No optical measurements of the inclination have been reported, hence we do not use it in our calculations.

\cite{daag} monitored the evolution of this source during outburst for three months at different wavelengths, and showed that the transition from the hard state to the soft state is around day 103 \cite[i.e. $\sim$ MJD 56180, see Fig 2 of][]{daag}. Multi-wavelength light curves and color evolution \cite[see Fig 3 of][]{daag} showed an unusual drop followed by a rise in all bands just before this state transition, with no color change. So the flux drop we calculate is after that, between day ~103 and ~109 (MJD ~56180 to ~56186), when the color change occured, indicating a decrease in jet emission. We find that the IR emission drop over the hard to soft transition is 0.41$\pm$0.28 mag. \\

\textit{Final sample :} Our final sample consists of only those 9 sources for which we have reliable constrains for all of the required parameters. We remove H 1743-322 from our analysis, as the masses and the orbital period are not known for this source (although it is included in Fig. 3). We exclude XTE J1752-223 from our final sample, as it has only upper limits for both the inclination angle and the orbital period, while the mass of the companion star is completely unknown. We also remove Swift J1910.2-0546 and MAXI J1535-571 as no optical and radio measurements of the inclination have been reported in the literature. And finally, as there are various model-dependent values of the inclinations reported for GX 339-4 in the literature (with inclination values in disagreement with each other by more than 30 degrees), we also exclude this source from our analysis, although we discuss GX 339-4 later in more detail. In Table 1, we list all of the BHXBs with measured IR excess, and we identify the excluded sources in italics.

\section{MODEL}
\label{sec:res}

\begin{figure}
\center
\includegraphics[height=0.4\textwidth]{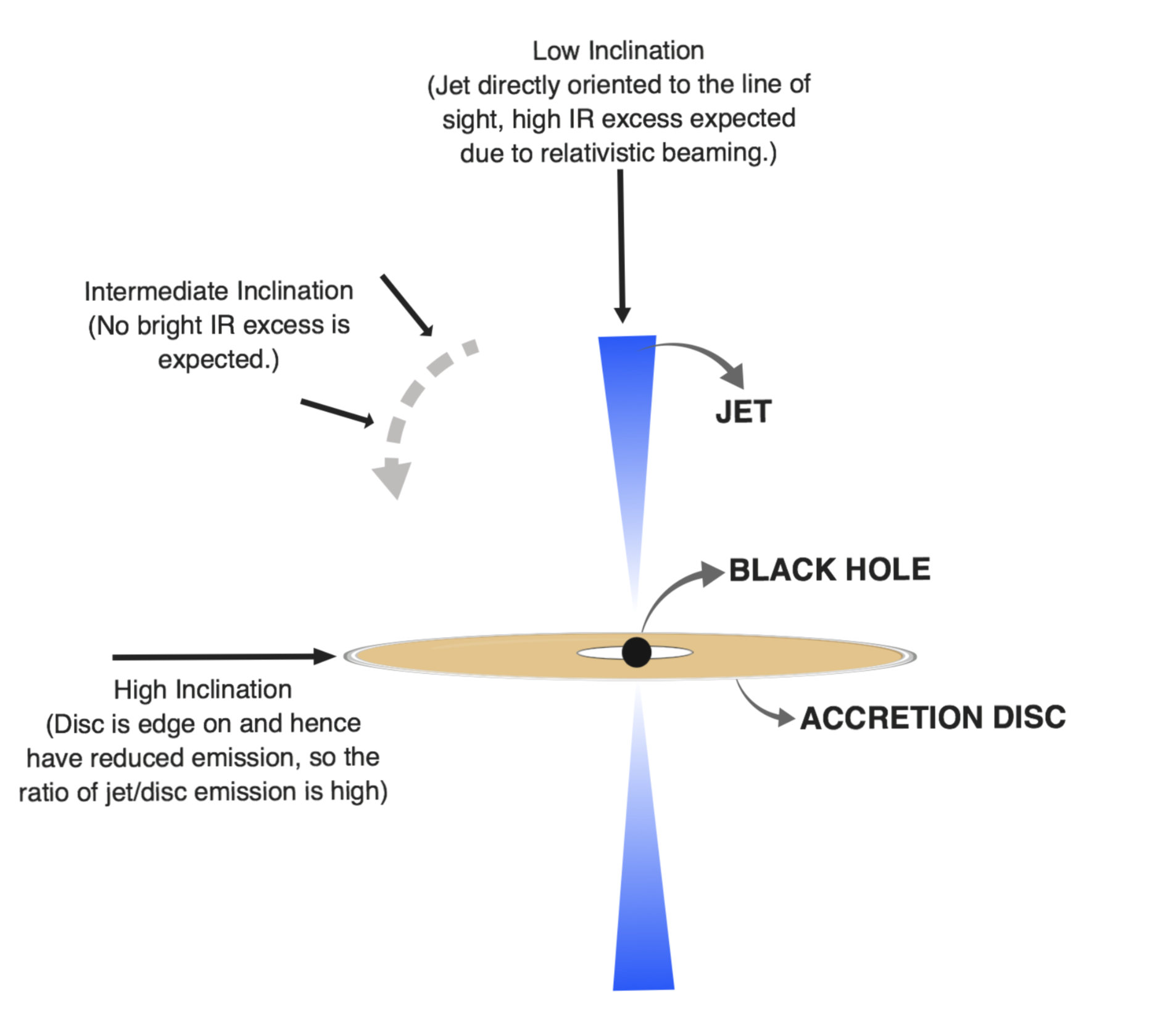}
\caption{Schematic diagram to qualitatively explain the expected IR excess observed for different BHXB sources, depending on their inclinations.}
\end{figure}

\subsection{Qualitative theoretical prediction}

Theoretically, the jet luminosity should correlate with the inclination angle if the emission is subject to relativistic beaming. Indeed, it has been shown that some of the sources with the brightest IR excesses compared to their discs are systems with a low inclination angle (e.g. 4U 1543-47 and MAXI J1836-194; see table 1 and Fig. 3).

For high inclination systems, the disc is almost edge-on, which reduces the disc emission but not the jet emission (if the jet is not highly beamed). This can lead to relatively bright IR excesses in high inclination systems. For intermediate inclination angles ($\sim$30 - 60 deg), no bright IR excess is expected for any Lorentz factor, and indeed all of the BHXBs with inclination angles within this range do not have prominent IR excesses (see Fig 3). When the inclination is low, the jet is oriented directly towards the line-of-sight, and a prominent IR excess is expected due to relativistic beaming (see Fig 2).

\begin{figure*}
\center
\includegraphics[height=0.62\textwidth]{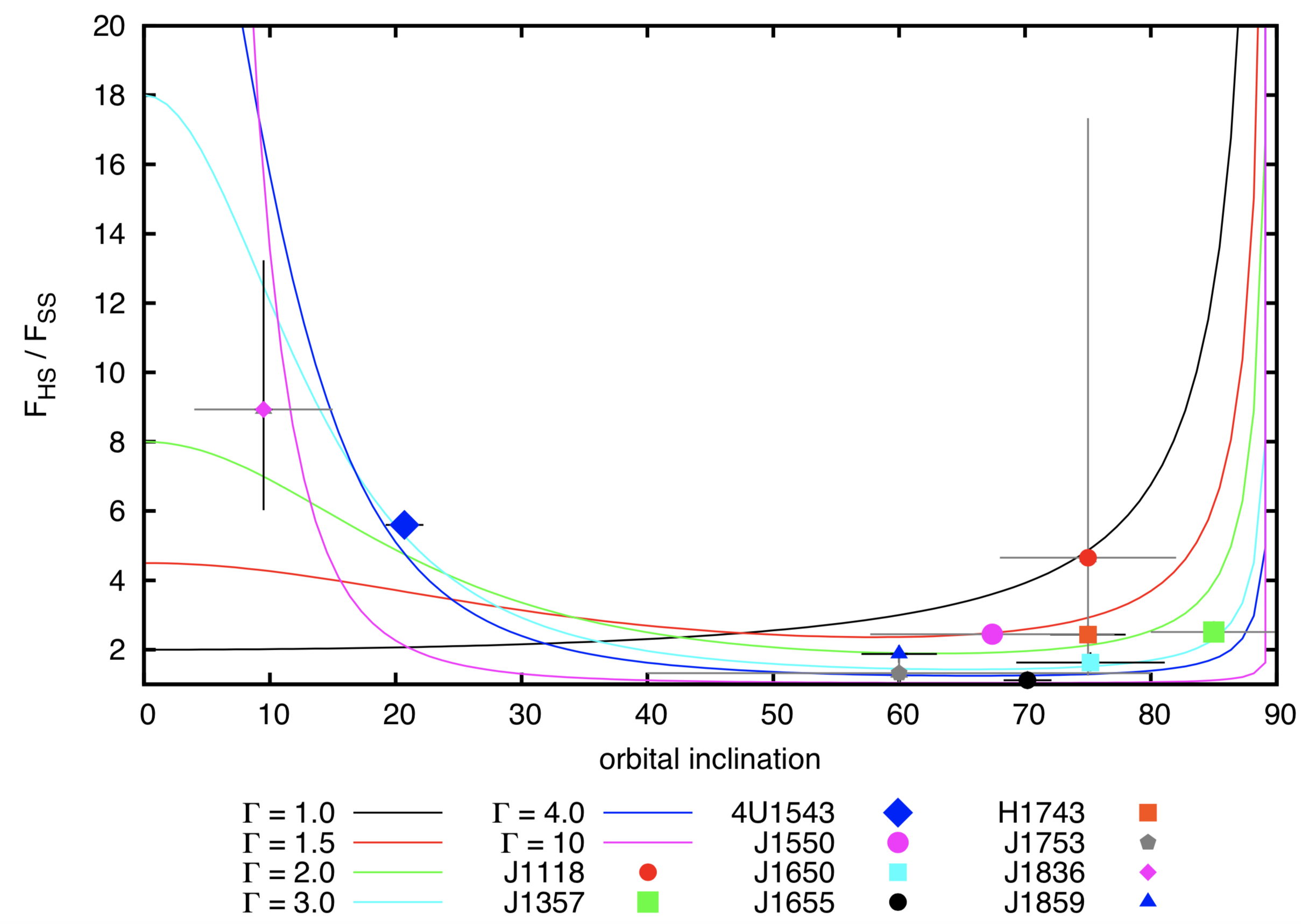}
\caption{The different lines show the flux change when the jet switches on and off (shown as a ratio of IR emission in the hard state vs the soft state), vs the orbital inclination for different Lorentz factors. The points with error-bars include all the BHXBs with well-constrained measurements of inclination and reliable estimates of IR excess. The error-bars represent the range of values in the case of a uniform distribution, and the standard deviation for a gaussian distribution, as specified in Table 1.}
\end{figure*}

\subsection{Model prediction of the IR excess}

When a BHXB is in the soft state, the jet is quenched and hence any NIR flux observed can be assumed to originate from the accretion disc. When the source transits to the hard state, the jet reappears and we see an excess of flux in the NIR wavelength range. If the appearance/disappearance of the NIR excess observed in BHXBs during state transition is occurring because the jet switches on/off, then we can use a simple analytical model to predict the relative flux excess expected for each source. To do this, we calculate the projected area of the accretion disc (to estimate the relative disc emission) and the Doppler beaming as a function of the Lorentz factor (to infer the jet emission). 

The projected area of the accretion disc is calculated depending on the size of the disc and the inclination angle $i$. Assuming that the disc is irradiated by a point source, the disc reprocessing scales as $R^{2/7}$, where $R$ is the radius of the outer disc \citep[see the discussion along with equations 5.94 and 5.95 in][and references therein]{book2002}. So projected (observed) disc reprocessing $\propto R^{2/7} \cos(i)$, where $R$ is proportional to the orbital separation, and the orbital separation is proportional to $(M_{\rm BH} + M_{\rm CS})^{1/3} P_{\rm orb}^{2/3}$ \citep[e.g.][]{van1994,r06}. Hence, the projected (observed) disc reprocessing obeys 

\begin{equation}
D_{\rm pr} \propto (M_{\rm BH} + M_{\rm CS})^{2/21}  \, P_{\rm orb}^{4/21} \, \cos(i) .
\end{equation}

Hence, the IR disc emission is $F_{\rm disc} = k_1 D_{\rm pr}$, where $k_1$ is an unknown constant that depends on e.g. the disc reprocessing efficiency.

The Doppler factor for jet emission $\Delta_{\rm jet}$ represents the observed radio luminosity of the jet relative to its intrinsic (rest-frame) radio luminosity. This relation can be described as $S_{\rm jet,obs} = \Delta_{\rm jet} \times \, S_{\rm jet,intr}$. $\Delta_{\rm jet}$ is calculated following the same method as in \cite{g03} via

\begin{equation}
\Delta_{\rm jet} = \frac{(\delta_{\rm rec})^2+(\delta_{\rm app})^2}{2} ,
\end{equation}

where the receding and approaching factors $\delta_{\rm rec/app} = \Gamma^{-1} (1 \pm \beta \cos i)^{-1}$; $\beta = v/c$ is the bulk velocity of the radio-emitting material relative to the speed of light, and $\Gamma = (1-\beta^{2})^{-1/2}$ is the corresponding bulk Lorentz factor. 

In the soft state, when we do not expect a jet, the IR emission should entirely consist of the disc emission (proportional to $ D_{\rm pr}$). On the other hand, both the disc and the jet should contribute towards the IR emission in the hard state (where the excess IR flux is proportional to $\Delta_{\rm jet}$; $F_{\rm jet} = k_2 \Delta_{\rm jet}$). Hence, for each of the sources, we can use the calculated projected area of the disc, and the Doppler factor of the jet, to estimate the predicted magnitude change $\Delta m_{\rm IR,pred}$ that we expect over state transition. This quantity is obtained via

\begin{equation}
\begin{split}
\Delta m_{\rm IR,pred} & = 2.5 \log_{10} \left[ \frac{k_1 D_{\rm pr} + k_2 \Delta_{\rm jet}}{k_1 D_{\rm pr}} \right] \\ 
& = 2.5 \log_{10}  \left[ 1+ C\frac{\Delta_{\rm jet}}{D_{\rm pr}}  \right] ,
\end{split}
\end{equation}

where $C = k_{2}/k_{1}$ is the ratio of the two unknown constants, and depends on the disc and jet radiative efficiencies. In our model, we are essentially testing if the disc size and inclination angle can be used to predict the jet/disc flux ratio in the hard state. For a source at a given disc size and inclination angle, we can predict the jet/disc flux ratio. This requires $C$ to have the same value for all sources.Theoretically, it is known for reprocessing of the disc, that the optical luminosity \citep[and hence also the IR luminosity, see][]{r06} is proportional to the X-ray luminosity as L$_{\rm Opt} \propto$  L$_{\rm IR} \propto$ L$_{\rm X}^{0.5}$ \citep{van1994}. So we expect the relation $k_1 \propto$ L$_{\rm X}^{0.5}$. For the jet in the hard state, assuming a flat spectrum from radio to IR wavelengths, and using the fundamental plane, we get L$_{\rm R} \propto$  L$_{\rm IR} \propto$ L$_{\rm X}^{0.7}$. Hence we can assume, $k_2 \propto$ L$_{\rm X}^{0.7}$, and roughly $C = k_{2}/k_{1} \propto$ L$_{\rm X}^{0.2}$. Therefore, C can be considered roughly constant, and to be having a very weak dependency on both the X-ray luminosity and mass accretion rate. Many observational studies have also confirmed the expected theoretical correlations, mostly following the relation L$_{\rm Opt/IR} \propto$ L$_{\rm X}^{\sim0.6}$ \citep[eg.][]{ho05,cor,ber,vin}. So the dependency of the constant $C$ on L$_{\rm X}$ is even shallower ( $C \propto$ L$_{\rm X}^{0.1}$) when using the observational relations.

Although there are caveats in this assumption, we do not think they should change the dependency much. One
caveat is the observed jet break in the spectrum that is often seen in the IR band, which mostly lies
in the optically thin part of the synchrotron spectrum. If the jet break frequency also scales with the X-ray luminosity,
then the jet IR emission will no longer be proportional to the jet radio emission. But as
shown in \cite{dave13}, there appears to be no strong relation between the jet break frequency and
L$_{\rm X}$ in the hard state. Hence we can expect L$_{\rm R}$ to be proportional to L$_{\rm IR}$, which is also seen
observationally in \cite{r06}. Another caveat is that the relation L$_{\rm R} \propto$ L$_{\rm X}^{0.7}$ does not hold observationally in all BHXBs. The radio-faint sources seem to have a shallower relation at low L$_{\rm X}$ and a steeper relation at high L$_{\rm X}$. But, it has been found that the radio-faint sources are also IR-faint, so  L$_{\rm R} \propto$  L$_{\rm IR}$ from the jet should still hold. So we do not expect these caveats to significantly change the assumption of weak dependency of the constant $C$ on L$_{\rm X}$ and mass accretion rate. In any case, it is important to note that if other parameters play a role, such as different reprocessing efficiencies in different discs, or different jet properties resulting in different jet fluxes for the same inclination and Lorentz factor, then we expect our model to provide a poor description of the data.\\

Fig 3 shows the theoretical prediction of the ratio of IR emission in the hard state vs the soft state (i.e. $10^{0.4\Delta m_{\rm IR,pred}}$) as a function of the inclination angle for various Lorentz factor values, assuming the disc radius and C to be the same for all ($C$=1 and $R$=1, i.e. $D_{\rm pr}=$cos$(i)$). As expected, we see that the IR excess is high for both very low and very high inclination values, while the intermediate inclination range shows very little IR excess for all Lorentz factors. We also plot all of the BHXBs for which IR excess and inclination angles are well-constrained. It is encouraging to note that all the BHXB sources occupy the region theoretically predicted by our model. 

\subsection{Model uncertainties}

The uncertainty on the Doppler factor $\Delta_{\rm jet}$ is dominated by the uncertainty related to the inclination of the source. On the other hand, the uncertainty on the projected disc area depends on the mass of the compact object, the mass of the companion star, the orbital period and the inclination angle. The dominant source of uncertainty in calculating the expected IR flux change is the inclination angle, because the projected disc area is proportional to the $\frac{2}{21}$th power of the sum of its component masses, and the $\frac{4}{21}$th power of the orbital period (see Eq. 1). Hence it is essential to use only those sources for the analysis which have a reliable estimate of the inclination. As described in Section 2, we only use inclination angles derived either from optical measurements or through radio observations, to maintain uniformity and reliability. \\

Since the uncertainties involved in the masses of the compact object and its companion star do not contribute much to the IR magnitude change uncertainty, this gives us the possibility of using a standard mass range for compact object and companion stars in BHXBs if these parameters are not well-known for a source. For example, when we have just a lower limit for the mass of the compact object, we place a conservative upper limit of 25$M_{\odot}$ in our Bayesian analysis. We also assume a conservative lower limit of 2$M_{\odot}$ for the mass of the compact object (below which it would be a neutron star). Similarly, we place a conservative lower limit of 0.08 $M_{\odot}$ on the mass of the companion star (below which is the brown dwarf regime). Finally, we also enforce a lower limit of 2 hours on the orbital period, since no dynamically confirmed BHXB system is known to have an orbital period of less than 2 hours. 

\section{Jet Lorentz Factor Estimation}

\begin{figure}
\center
\includegraphics[height=0.28\textwidth]{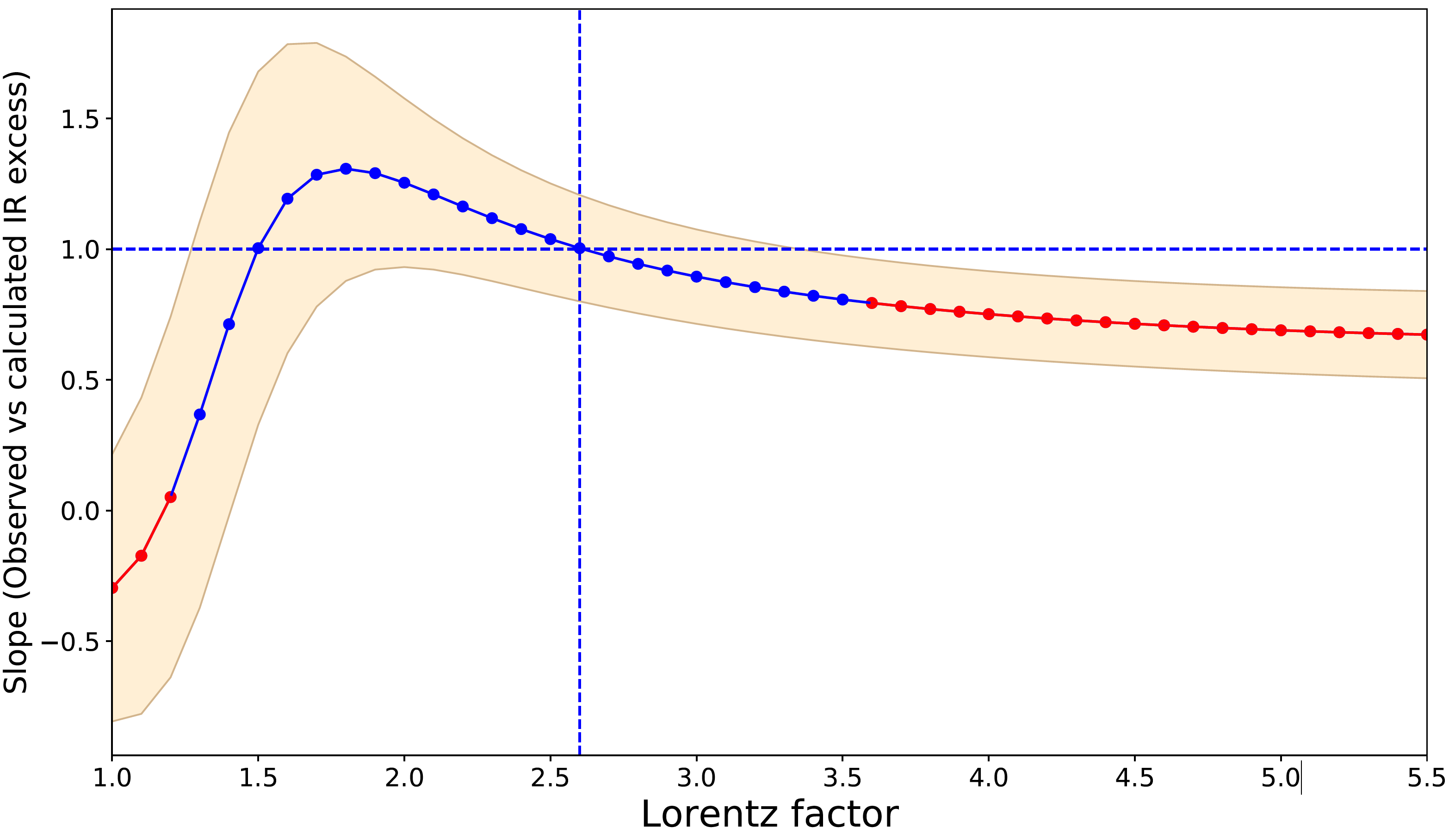}
\caption{Best Lorentz factor values (when C=1) by looking at the slopes. A slope of 1 (in the Y-axis) implies that the distribution of observed IR excess matches with the predicted one, for that particular Lorentz factor. The shaded region represents the one-sigma error on the value of slope. We see that only the Lorentz factors in the range of 1.3-3.5 (shown here in blue points) are in agreement with having a slope of 1 within the errors.}
\end{figure}

We aim to constrain for the first time the jet Lorentz factors for the nine BHXBs with good data from Table~1. We will do this by modelling the observed IR magnitude change
$\Delta m_{\rm IR,obs}$ for each BHXB using the prior constraints on inclinations, component masses, and orbital periods tabulated in Table~1, along with our simple analytical model
specified in Section~3.2 (namely Equations 1, 2 and 4).

\begin{figure}
\center
\includegraphics[height=0.34\textwidth]{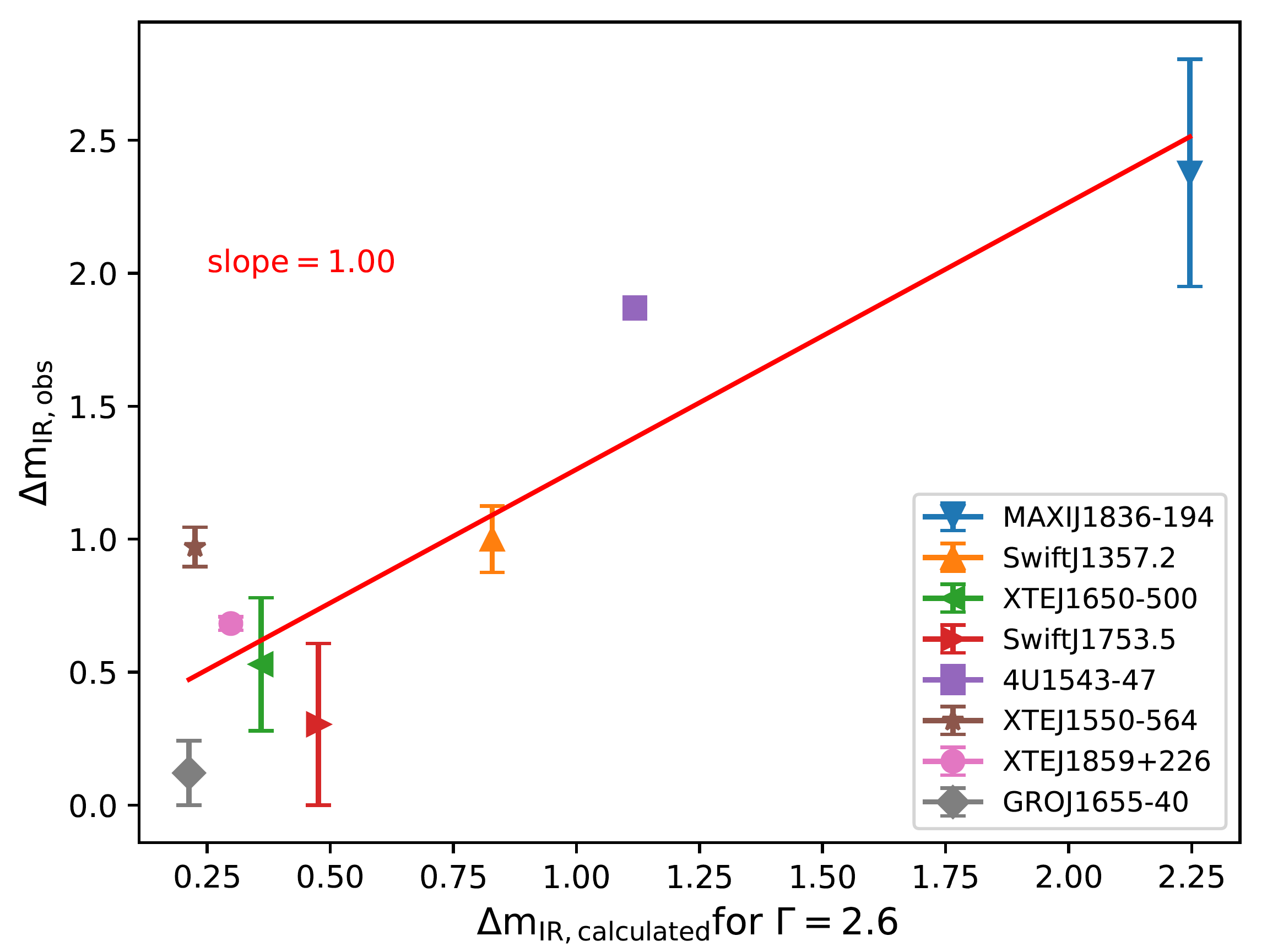}
\includegraphics[height=0.34\textwidth]{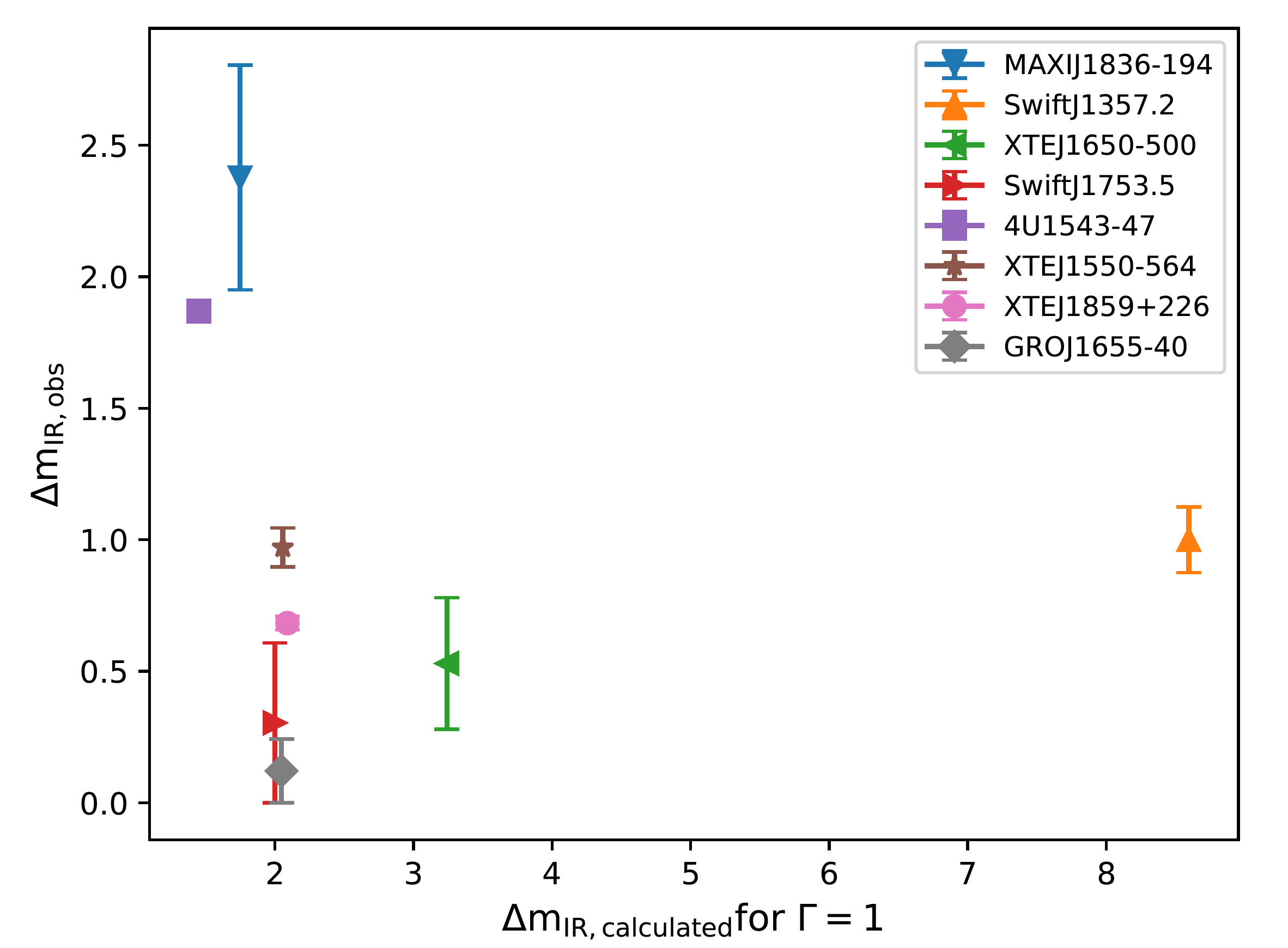}
\caption{\textit{Upper panel:} The predicted vs observed NIR magnitude change when the jet switches on and/or off, assuming the best-fit Lorentz factor of 2.6. \textit{Lower panel:} No correlation is visible between the predicted observed IR magnitude change for a Lorentz factor = 1.}
\end{figure}

As a first test, we initially assume a common Lorentz factor for all BHXBs, to investigate if relativistic beaming is likely to be a realistic cause of the differing IR excess amplitudes between sources. For this, we calculate the predicted IR excess ($\Delta m_{\rm IR,pred}$, using Eq 1, 2 and 4) for each of these sources using different Lorentz Factors, and assuming the constant to be $C$=1. In this test, we do not include XTE J1118+480 as there are two model-dependent values of observed IR excess for this source, which differs from each other by more than 2.5 mag. A preliminary way of estimating which Lorentz factor range best reproduces the observed IR excess seen in BHXBs, is to just check the slope between the observed and predicted values of IR magnitude change during state transition. A slope of 1 will automatically mean that the distribution of observed IR excess matches with the calculated one for that particular Lorentz factor. As shown in Fig 4, we see that only for the Lorentz factors in the range of 1.3 - 3.5, the predicted IR excesses are in agreement with the observed ones. This approach suggests a Lorentz factor value of 2.6, for which the observed and predicted flux ratios are highly correlated with a slope of 1. The slope systematically deviates from 1 and the correlation becomes weaker as we move to other Lorentz factors (see Fig 4).

For the Lorentz factor of 2.6, we show the predicted versus the observed magnitude change in Fig. 5. There is clearly a strong correlation (Pearson r-coefficient = 0.9, p-value = 0.001), indicating that it is very likely that the Lorentz factor plays an important role in the amplitude of the IR excess in BHXBs. For Lorentz factor = 1, i.e. non-beamed emission, we find that there is no correlation at all (Pearson r-coefficient = -0.2, p-value = 0.461). This strongly supports the IR emission in the hard state to be outflowing and beamed, with a likely Lorentz factor of 1.3 - 3.5 (when C=1). But it is highly unlikely that all the BHXBs will have just one Lorentz factor. For a proper analysis, we turn to Bayesian statistics.

\subsection{Hierarchical Bayesian Model}

For a detailed study, we adopt a Bayesian framework in which to analyse our data. Firstly, let us employ a subscript $j$ to denote a specific object, with $j$ running from 1 to 9. Then,
for the $j$th BHXB, we have an observed datum  $\Delta m_{{\rm IR,obs},j}$, and five BHXB-specific model parameters $i_{j}$, $M_{{\rm BH} ,j}$, $M_{{\rm CS},j}$, $P_{{\rm orb},j}$, 
and $\Gamma_{j}$. The parameter $C$ in our model applies to all BHXBs as discussed in Section~3.2. The parameters $i_{j}$, $M_{{\rm BH},j}$, $M_{{\rm CS},j}$, 
and $P_{{\rm orb},j}$ all have prior constraints, whereas the parameters $\Gamma_{j}$ do not. We therefore assume that the parameters $\Gamma_{j}$ are drawn from a parent distribution common to BHXBs, parametrised by a single parameter $\alpha$ (see later in this section). In total we have 9 data points, 47 parameters, 
and various prior constraints.\\

For notational convenience, let:
\begin{equation}
\mathbf{D} \equiv \left\{ \Delta m_{{\rm IR,obs},j} \right\}_{j=1}^{j=9}
\end{equation}
\begin{equation}
\bm{\theta}_{j} = \left( i_{j}, M_{{\rm BH},j}, M_{{\rm CS},j}, P_{{\rm orb},j} \right)
\end{equation}
\begin{equation}
\mathbf{\Theta} = \left( \bm{\theta}_{1}, \dots, \bm{\theta}_{9}, C \right)
\end{equation}
\begin{equation}
\mathbf{G} = \left( \Gamma_{1}, \dots, \Gamma_{9} \right)
\end{equation}

Using $\mathcal{M}$ to represent the entirety of our model and its assumptions, we may write the posterior probability distribution over all of our parameters as:
\begin{equation}
\begin{array}{l}
P \left( \mathbf{\Theta}, \mathbf{G}, \alpha \, | \, \mathbf{D}, \mathcal{M} \right) \\
\,\,\,\,\,\, \propto P \left( \mathbf{D} \, | \, \mathbf{\Theta}, \mathbf{G}, \mathcal{M} \right) \, 
                     P \left( \mathbf{\Theta} \, | \, \mathcal{M} \right) \,
                     P \left( \mathbf{G} \, | \, \alpha, \mathcal{M} \right) \,
                     P \left( \alpha \, | \, \mathcal{M} \right)
\end{array}
\end{equation}
where the prior probability distribution for the subset of model parameters $\mathbf{\Theta}$ is given by:
\begin{equation}
\begin{array}{l}
P \left( \mathbf{\Theta} \, | \, \mathcal{M} \right) = P \left( C \, | \, \mathcal{M} \right) \,
                                                       \prod_{j = 1}^{9} \left[ \, P \left( i_{j} \, | \, \mathcal{M} \right) \, P \left( M_{{\rm BH},j} \right) \, | \, \mathcal{M} \right) \\
\,\,\,\,\,\,\,\,\,\,\,\,\,\,\,\,\,\,\,\,\,\,\,\,\,\,\,\,\,\,\,\,\,\,\,\,\,\,\,\,\, \,\,\,\,\,\, \,\,\,\,\,\,\,\,\,\,\,\,\,\,\,\, \times \, P \left( M_{{\rm CS},j} \, | \, \mathcal{M} \right) \, P \left( P_{{\rm orb},j} \, | \, \mathcal{M} \right) ]
\end{array}
\end{equation}
These expressions have been derived using Bayes Theorem, the definition of conditional probability, and by assuming independence of the prior probabilities.

The prior constraints (PDFs) on the parameters $i_{j}$, $M_{{\rm BH},j}$, $M_{{\rm CS},j}$, and $P_{{\rm orb},j}$ that feature in Equation 10 are listed in Table~1.
If the entry for a parameter in Table 1 is given as a range, then the prior PDF for the parameter is assumed to be a uniform distribution over the
range (except for inclination where the prior PDF is assumed to be uniform in $\cos(i_{j})$ over the range to satisfy isotropic orientation). If the entry
for a parameter in Table~1 is given in the form $\mu \pm \sigma$, then the prior PDF for the parameter is assumed to be a Gaussian distribution with
mean $\mu$ and standard deviation $\sigma$, and with hard cutoffs at 0$^{\circ}$ and 90$^{\circ}$ for $i_{j}$, 2$M_{\odot}$ and 25$M_{\odot}$ for $M_{{\rm BH},j}$,
0.08$M_{\odot}$ for $M_{{\rm CS},j}$, and 2~h for $P_{{\rm orb},j}$ (see Section~3.3). Finally, if the entry for a parameter in Table~1 is given as a lower/upper
limit, then the prior PDF for the parameter is assumed to be a uniform distribution over the range up/down to the hard cutoffs previously mentioned.
The prior PDF $P(C \, | \, \mathcal{M})$ is chosen to be uniform in $\ln C$ with unrestricted range so as to limit $C$ to strictly
positive (physical) values while assuming that all orders of magnitude for $C$ are equally likely \textit{a priori}.
These priors effectively serve as a regularisation of the 37 model parameters in $\mathbf{\Theta}$ in the Bayesian inference problem.

\begin{figure*}
\center
\includegraphics[height=0.90\textwidth]{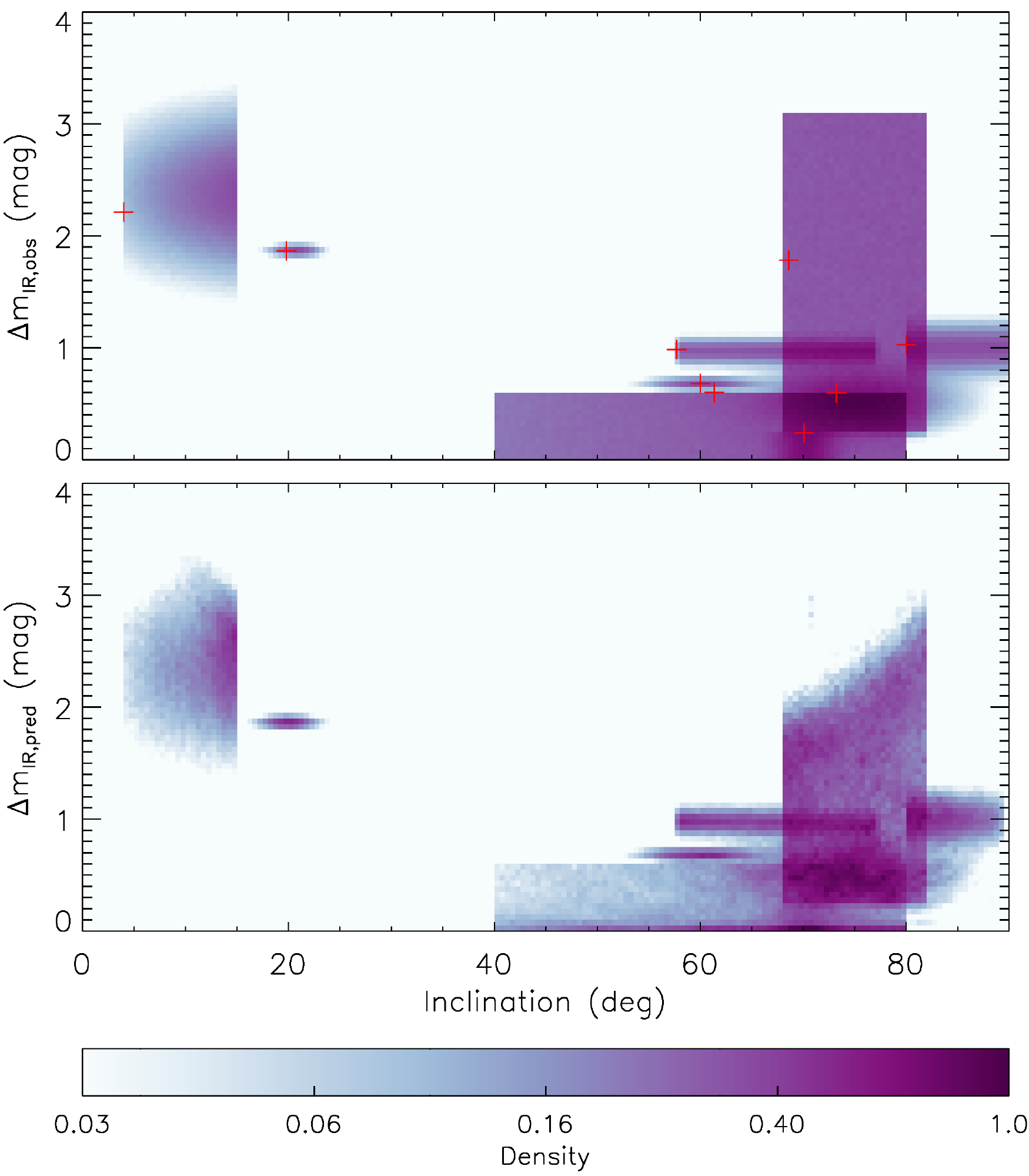}
\caption{\textit{Upper panel:} Plots of surfaces, for each of the nine BHXBs, that are proportional to the prior probability density as a function of the inclination $i$ ($x$-axis) and observed IR excess $\Delta m_{{\rm IR,obs}}$ ($y$-axis). Red crosses show the best-fit model values for each source. The relevant probability densities for each BHXB are defined in Table~1. \textit{Lower panel:} Two-dimensional histograms (normalised by peak), for each of the nine BHXBs, of posterior inferred inclination $i$ ($x$-axis) and predicted IR excess $\Delta m_{{\rm IR,pred}}$ ($y$-axis) constructed using the 2$\times$10$^{6}$ MCMC samples  (see Section~4.2). \textit{Bottom:} Colour-intensity bar (logarithmic scale) for the density surfaces in each panel.
         \label{fig:pdm}}
\end{figure*}

The next ingredient required in our hierarchical Bayesian model is the definition of the parent distribution for the jet Lorentz factors of BHXBs.
For AGN jets, many studies find bulk Lorentz factor distributions in the form of a power-law \citep[$N(\Gamma) \propto \Gamma^{\alpha}$ with $\alpha<-1$;][etc. See Section~5.3 for a detailed discussion]{pu92,lm97,s16}. Therefore, we adopt the following parent
PDF for each parameter $\Gamma_{j}$:
\begin{equation}
P \left( \Gamma_{j} \, | \, \alpha, \mathcal{M} \right) =
\begin{cases}
0                                    & \text{for $\Gamma_{j} < 1$}    \\
-(\alpha + 1) \, \Gamma_{j}^{\alpha} & \text{for $\Gamma_{j} \geq 1$} \\
\end{cases}
\end{equation}
and, assuming independence for the $\Gamma_{j}$, we may write:
\begin{equation}
P \left( \mathbf{G} \, | \, \alpha, \mathcal{M} \right) = \prod_{j = 1}^{9} P \left( \Gamma_{j} \, | \, \alpha, \mathcal{M} \right)
\end{equation}
For the (hyper)-prior PDF $P \left( \alpha \, | \, \mathcal{M} \right)$, we assume a uniform distribution over the range negative infinity to $-$1.

The final ingredient required in our hierarchical Bayesian model is an expression for the likelihood function of our data $\mathbf{D}$,
represented as $P \left( \mathbf{D} \, | \, \mathbf{\Theta}, \mathbf{G}, \mathcal{M} \right)$
. We assume that all data points are independently observed, and we may therefore write:
\begin{equation}
P \left( \mathbf{D} \, | \, \mathbf{\Theta}, \mathbf{G}, \mathcal{M} \right) = \prod_{j = 1}^{9} P \left( \Delta m_{{\rm IR,obs},j} \, | \, \bm{\theta}_{j}, \Gamma_{j}, C, \mathcal{M} \right)
\end{equation}
The individual data-point likelihoods may be computed by adopting a noise model as part of our full model $\mathcal{M}$. We have:
\begin{equation}
\Delta m_{{\rm IR,obs},j} = \Delta m_{{\rm IR,pred},j} + \epsilon_{j}
\end{equation}
where $\epsilon_{j}$ is a noise contribution, and $\Delta m_{{\rm IR,pred},j}$ is computed from the parameters
$i_{j}$, $M_{{\rm BH},j}$, $M_{{\rm CS},j}$, $P_{{\rm orb},j}$, $\Gamma_{j}$, and $C$ (using equations 1, 2 and 4).
In Table~1, the IR excess is listed either as a range $a$ to $b$
or as two numbers $a \pm b$. For a BHXB with the IR excess listed as a range, we set:
\begin{equation}
\Delta m_{{\rm IR,obs},j} = \frac{1}{2} (a + b)
\end{equation}
\begin{equation}
P \left( \epsilon_{j} \, | \, \mathcal{M} \right) = U \left( \epsilon_{j} \, | \, -(b - a)/2, (b - a)/2 \right)
\end{equation}
where $U(x \, | \, u,v)$ represents a uniform distribution with lower and upper limits $u$ and $v$, respectively.
Otherwise, we set:
\begin{equation}
\Delta m_{{\rm IR,obs},j} = a
\end{equation}
\begin{equation}
P \left( \epsilon_{j} \, | \, \mathcal{M} \right) = N \left( \epsilon_{j} \, | \, 0, b \right)
\end{equation}
where $N(x \, | \, u, v)$ represents a Gaussian distribution with mean $u$ and standard deviation $v$.
Putting this together, we have:
\begin{equation}
P \left( \Delta m_{{\rm IR,obs},j} \, | \, \bm{\theta}_{j}, \Gamma_{j}, C, \mathcal{M} \right) = P \left( \epsilon_{j} \, | \, \mathcal{M} \right)
\end{equation}

In the top panel of Figure~6, for each BHXB, we plot surfaces that are proportional to the prior probability density as a function of inclination $i$ ($x$-axis) and observed IR excess $\Delta m_{{\rm IR,obs}}$ ($y$-axis). The surfaces are defined by the prior PDFs for the inclinations,
and the observed $\Delta m_{{\rm IR,obs}}$ values along with the adopted noise model, that are listed in Table~1. A good model for the data $\mathbf{D}$
should be able to reproduce the peak densities for $\Delta m_{{\rm IR,obs}}$ without yielding inferred inclinations that are too far away from the peaks of their
corresponding prior PDFs.

The inspiration and guidance for the above development of a hierarchical Bayesian model has been taken from studying the papers by
\cite{hogg} and \cite{kelly}, although there are many examples of usage of this modelling
technique in the recent astronomical literature.

\subsection{Parameter Inference}

\begin{table*}
\centering
\begin{tabular}{ | l | r | r | r | r | r | r | }
\hline
Name               & Inclination ($^\circ$) & $M_{\rm BH}$ ($M_{\odot}$) & $M_{\rm CS}$ ($M_{\odot}$) & $P_{\rm orb}$ (hrs) & $\Gamma$ & $\Delta m_{{\rm IR,pred}}$ \\
\hline
XTE J1118+480      & 68.6                   & 7.31                       & 0.190                      & 4.0784143           & 1.00     & 1.782                      \\
Swift J1357.2-0933 & 80.0                   & 25.00                      & 0.402                      & 2.82                & 2.39     & 1.028                      \\
4U 1543-47         & 19.80                  & 9.40                       & 2.453                      & 26.793771           & 2.47     & 1.865                      \\
XTE J1550-564      & 57.70                  & 9.12                       & 0.302                      & 37.0088025          & 1.36     & 0.986                      \\
XTE J1650-500      & 73.23                  & 4.91                       & 2.360                      & 7.690               & 2.82     & 0.599                      \\
GRO J1655-40       & 70.04                  & 5.41                       & 1.471                      & 62.92580            & 3.88     & 0.240                      \\
Swift J1753.5-0127 & 61.37                  & 25.00                      & 0.250                      & 3.240               & 2.74     & 0.600                      \\
MAXI J1836-194     & 4.0                    & 14.38                      & 0.638                      & 4.90                & 1.72     & 2.210                      \\
XTE J1859+226      & 60.0                   & 11.57                      & 5.410                      & 6.580               & 2.37     & 0.682                      \\
\hline
$C$                & 2.410                  & \,  &  \,  &  \,  &  \,  &  \,  \\
$\alpha$           & $-$2.316               & \, &  \,  &  \,  &  \,  &  \,   \\
\hline
\end{tabular}
\caption{Best-fit MAP parameter estimates for the hierarchical Bayesian model in Section~4.1. The two free parameters that are common to all the BHXBs, namely the constant ($C$) and the index of the parent power-law distribution of jet Lorentz factors ($\alpha$), are also reported at the end of the table.}
\end{table*}

\begin{table*}
\centering
\begin{tabular}{ | l | r | r  | }
\hline
Name               & Inclination ($^\circ$)                                                                      & $\Gamma$                                \\
\hline
XTE J1118+480      & 68.3, 70.2, \textbf{75.2}, 79.7, 81.7      & 1.02, 1.16, \textbf{1.76}, 3.44, 5.36   \\
Swift J1357.2-0933 & 80.2, 81.2, \textbf{83.8}, 87.0, 89.0      & 2.23, 2.66, \textbf{3.35}, 4.83, 8.25   \\
4U 1543-47         & 16.67, 18.38, \textbf{19.96}, 21.51, 23.13    & 1.45, 1.94, \textbf{2.67}, 3.79, 6.08   \\
XTE J1550-564      & 58.07, 60.32, \textbf{66.40}, 73.33, 76.57 & 1.11, 1.35, \textbf{1.61}, 1.90, 2.38   \\
XTE J1650-500      & 62.69, 68.53, \textbf{74.44}, 79.79, 84.95    & 2.14, 2.68, \textbf{3.54}, 4.99, 8.80   \\
GRO J1655-40       & 66.33, 68.21, \textbf{70.16}, 72.05, 73.95        & 3.90, 4.82, \textbf{8.21}, 25.13, 83.62 \\
Swift J1753.5-0127 & 41.80, 49.78, \textbf{63.21}, 74.25, 79.12                         & 2.96, 3.82, \textbf{6.74}, 19.25, 59.59 \\
MAXI J1836-194     & 4.8, 7.8, \textbf{11.8}, 14.1, 14.9        & 1.17, 1.45, \textbf{1.94}, 6.00, 22.90  \\
XTE J1859+226      & 53.83, 56.80, \textbf{59.94}, 62.92, 65.73     & 2.01, 2.26, \textbf{2.53}, 2.90, 3.58   \\
\hline
$C$                & 1.82, 2.18, \textbf{2.60}, 3.24, 4.74               &  \,  \\
$\alpha$           & $-$2.67, $-$2.22, \textbf{$-$1.88}, $-$1.61, $-$1.40 &   \,  \\
\hline
\end{tabular}
\caption{Inferred parameter estimates (bold) of the inclinations and Lorentz factor, and $\pm$1-sigma and $\pm$2-sigma credible regions, for the hierarchical Bayesian model in Section~4.1.
         Each entry is given as five numbers representing the 2.3, 15.9, 50 (bold), 84.1 and 97.7 percentiles of the MCMC samples. At the end, we also show the inferred values for the two free parameters that are common to all the BHXBs, namely the constant ($C$) and the index of the parent power-law distribution of jet Lorentz factors ($\alpha$). The complete table with all of the inferred parameters will be attached in the Appendix of the APJ published paper as Table 5.}
\end{table*}

We maximise the logarithm of the posterior probability density function in Equation 9 over the 47 parameters in our model by
using the Nelder-Mead simplex algorithm \citep{b1,b2} as implemented in the Python package {\tt scipy.optimize}. This yields a maximum \textit{a posteriori} (MAP) estimate of the best-fit parameters, which
we report in Table~3, and some of which are plotted in Figures~6 (upper panel), 7, and~8. Specifically, the best-fit parameters lead to a good fit
to the data $\mathbf{D}$ as demonstrated\footnote{Wherever a uniform distribution, or a hard limit, is employed in our model (i.e. in the prior PDFs and the noise model),
the maximum of the posterior PDF is likely to be found on the boundary of the allowed range, unless the allowed
range for a particular parameter is sufficiently large.
Hence the positioning of the red plus symbols at surface boundaries in the upper panel of Figure~6 is not to be unexpected.}
in the upper panel of Figure~6 (where the red plus symbols are the best-fit model values, while the coloured surfaces are the prior probability density functions). We find that the algorithm converges to the same MAP solution each time so long
as the initial values of the parameters that are constrained with uniform prior PDFs lie within the acceptable ranges (otherwise the algorithm fails to converge).

We sample the posterior PDF (Equation 9) using the MCMC ensemble sampler {\tt emcee} \citep{foreman}
with 1000 walkers initialised in a tight Gaussian ball around the MAP parameter estimates \citep{hforeman}.
Each walker executes a burn-in of 200 steps, and then iterates through
50000 subsequent steps of which the last 2000 steps are recorded. Due to the relatively high dimensionality of the model, we found that it was
necessary to set the proposal scale parameter to a non-default value of $a~=~1.15$ so as to achieve a mean acceptance fraction for each walker of $\sim$0.25.
In the lower panel of Figure~6, we use the 2$\times$10$^{6}$ MCMC samples to plot two-dimensional histograms, for each BHXB, of
posterior inferred inclination $i$ ($x$-axis) and predicted IR excess $\Delta m_{{\rm IR,pred}}$ ($y$-axis). The MCMC samples clearly provide a good match to
the observed IR excess $\Delta m_{{\rm IR,obs}}$ values within the observational noise while further constraining some of the inferred inclinations
(compare the panels in Figure~6).

\begin{figure}
\center
\includegraphics[height=0.5\textwidth]{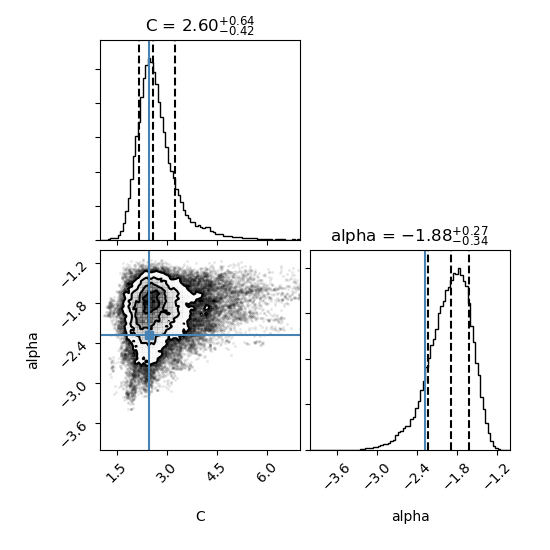}
\caption{\textit{Upper-left panel:} Histogram of the inferred parameter $C$ constructed using the MCMC samples. The best-fit MAP
                                    estimate is plotted with a vertical blue line, while the 15.9, 50 and 84.1 percentiles are plotted with three vertical dashed black lines.
         \textit{Lower-right panel:} Same as upper-left panel for the inferred parameter $\alpha$.
         \textit{Lower-left panel:} Two-dimensional histogram (normalised by peak) of the inferred parameters $C$ and $\alpha$, constructed using the MCMC samples.
                                    The 1-, 2-, and 3-sigma contours are also displayed. In the region outside of the 3-sigma contour, the individual samples are plotted. The
                                    best-fit MAP estimate for each parameter is also plotted with a blue line.
         \label{fig:corner}}
\end{figure}

\begin{figure*}
\center
\includegraphics[height=0.9\textwidth]{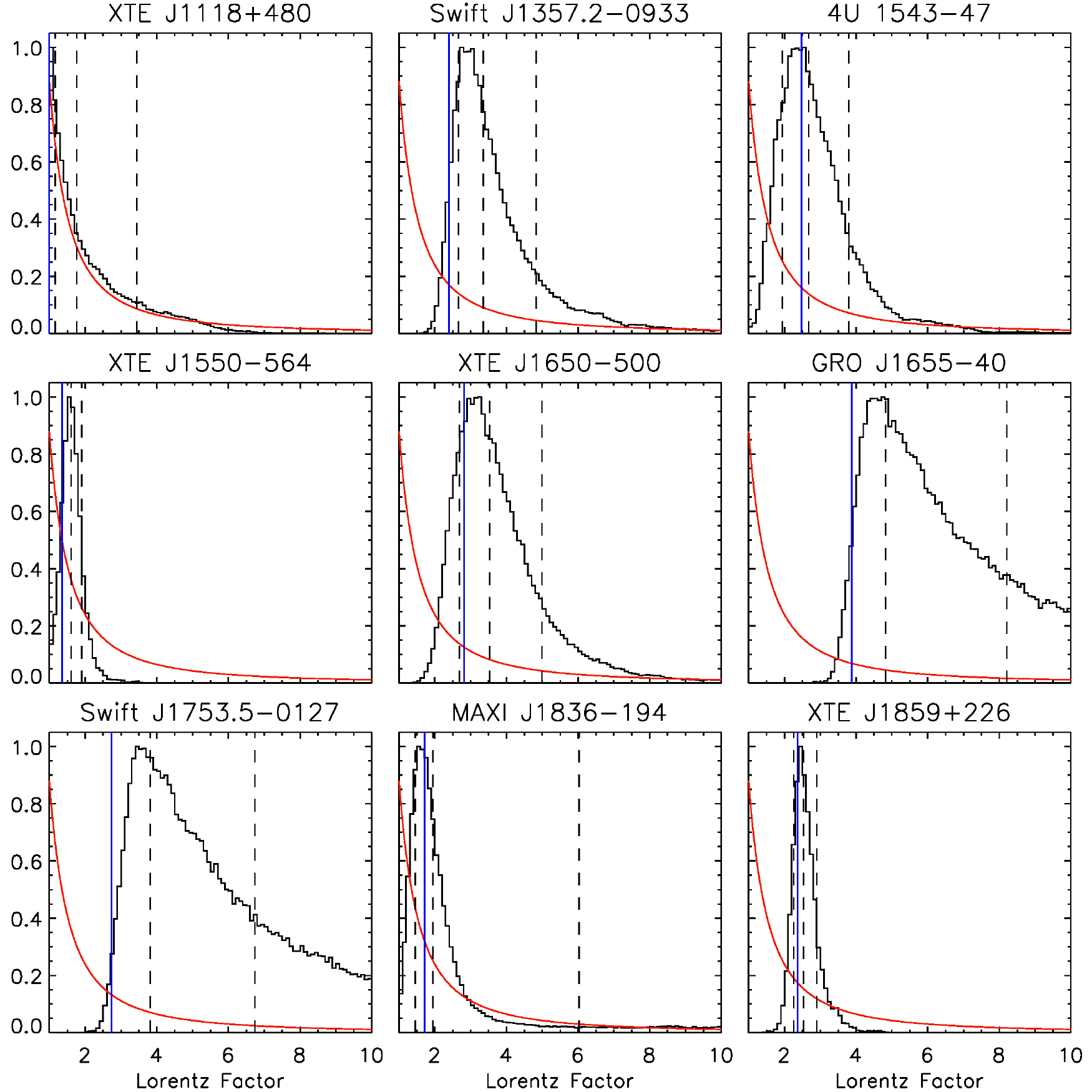}
\caption{Histograms (normalised by peak) of the inferred jet Lorentz factors $\Gamma_{j}$ for each BHXB, constructed using the MCMC samples.
         The best-fit MAP parameter estimates are plotted as vertical blue lines, and the 15.9, 50 and 84.1 percentiles are
         plotted as vertical dashed black lines. The red curve in each panel represents the inferred parent distribution for the jet Lorentz factors of BHXBs
         ($\alpha~=~-1.88$).
         \label{fig:Hist}}
\end{figure*}

In Table~4, we report the median parameter values for the MCMC samples as our inferred parameter estimates with $\pm$1-sigma and $\pm$2-sigma
credible regions computed using the 15.9 and 84.1 percentiles of the MCMC samples, and the 2.3 and 97.7 percentiles, respectively. It should be noted that these parameter estimates,
when taken together, do not represent a "best-fit" solution \citep[see][]{hforeman}. Our main result is that
$\alpha~=~-1.88^{+0.27}_{-0.34}$, which characterises the parent distribution for the jet Lorentz factors of BHXBs (see Section~5.2 for a discussion).
In Figure~7, we plot one-dimensional histograms of the model parameters $C$ and $\alpha$ along with a two-dimensional histogram
showing their covariance. The best-fit MAP parameter estimates ($\widehat{C}~\approx~2.410$ and $\widehat{\alpha}~\approx~-2.316$) are plotted with blue lines, while the
15.9, 50 and 84.1 percentiles are plotted with three vertical dashed black lines. The parameters $C$ and $\alpha$ are not degenerate, which
is fortunate considering their role as global parameters of our model.

The individual jet Lorentz factors $\Gamma_{j}$ for each BHXB are also of great interest. The inferred distributions are plotted as histograms (normalised
by peak) in Figure~8, with the best-fit MAP parameter estimates plotted as vertical blue lines, and the 15.9, 50 and 84.1 percentiles of the MCMC samples plotted as vertical dashed black lines.
The inferred parent distribution for the jet Lorentz factors of BHXBs is plotted as the red curve in each panel.
The posterior distribution of the Lorentz factor for XTE J1118+480 is essentially no different from the parent distribution, and it is therefore not usefully
constrained by the data in Table~1. Also, GRO~J1655-40 and Swift~J1753.5-0127 only have well-constrained lower limits to their jet Lorentz factors 
(see Table~4). However, the remaining BHXBs have useful posterior constraints on their jet Lorentz factors, especially
XTE~J1550-564, MAXI~J1836-194, and XTE~J1859+226 (see Section~5.2 for further discussion).\\

\section{Discussion}
\label{sec:disc}

\subsection{Lorentz Factor of BHXB jets from literature}

There is no clear consensus on the value of Lorentz factors expected in BHXB jets. Following \cite{mr94} it was suggested that BHXB jets would be significantly less relativistic than AGN jets, having $\Gamma\sim$ 2. \cite{fk2001} placed an upper limit of $\Gamma<$ 5, arguing that a value higher than that would probably destroy the observed correlation between radio and X-ray peak fluxes. Furthermore, \cite{g03} used the scatter in the radio/X-ray relation of BHXB jets to constrain the Lorentz factors of compact jets to $\Gamma<$ 2. 

There are also studies to estimate the Lorentz factor of compact jets in a few individual BHXBs. For example, \cite{cascas} used IR variability of GX 339-4 to constrain the Lorentz factor of the source to be $\Gamma>$ 2. \cite{tr2014} found an unusually steep radio/X-ray correlation for MAXI J1836-194 and argued that the Lorentz factor of this source needs to vary from $\Gamma\sim$1 at low X-ray luminosities to $\Gamma\sim$3-4 at high X-ray luminosities to produce the observed correlation. \cite{t2019} studied the time lags between the X-ray and radio bands for Cygnus X-1 and found a Lorentz factor value of 2.59$^{+0.79}_{-0.61}$.\\ 

On the other hand, for transient jets, \cite{fe04} predicted their Lorentz factors to be higher than that of compact jets, although \cite{miller06} did not find any significant observational evidence to support this claim. \cite{miller06} collected all available data on the opening angles of `ballistic' jets in BHXBs, calculated the Lorentz factors required to produce such small opening angles via the transverse relativistic Doppler effect, and found that the derived Lorentz factors have a mean value of $>$10 (which is much larger than for compact jets from other studies). 

But despite all these studies, it is still not clear what Lorentz factor values, or Lorentz factor distribution can be generally expected for jets in BHXBs. We use a novel approach to attempt to answer this question for compact jets. 

\subsection{Implication of our results for compact jets} 

We find that the parent distribution of the bulk Lorentz factors for jets in BHXBs is consistent with a power-law with the index $\alpha~=~-1.88^{+0.27}_{-0.34}$. This is very similar to their supermassive counterparts, where a bulk Lorentz factor distribution in the form of a power law is commonly found \citep[for example, a power law with $\alpha~=~-2.1\pm0.4$ for blazar jets, as obtained in][]{s16}.

We also estimate the individual Lorentz factors for each BHXB in our sample (see Table 4). We find that the Lorentz factor of most of the BHXB jets are quite low, for example $1.61^{+0.29}_{-0.26}$ for XTE~J1550-564, $1.94^{+4.06}_{-0.49}$ for MAXI~J1836-194, and $2.53^{+0.37}_{-0.27}$ for XTE~J1859+226. The Lorentz factors for Swift J1357.2-0933 was found to be $3.35^{+1.48}_{-0.69}$, 4U 1543-47 to be $2.67^{+1.12}_{-0.73}$ and XTE J1650-500 to be $3.54^{+1.45}_{-0.86}$. For two of the sources which did not have well-constrained values of observed IR excess (i.e. either had limits or model-dependent values), we can only provide limits on the bulk Lorentz factor. For example, for GRO~J1655-40 and Swift~J1753.5-0127 we can well-constrain only the lower limits of 3.90 and 2.96, respectively. This is because the IR excess for these two objects is consistent with being zero, which allows for very high Lorentz factors because they are high inclination sources. If more accurate IR excess measurements can be made, then their Lorentz factors could be better-constrained.

\subsection{Comparison of BHXB vs AGN jets}

The most common way to calculate the Lorentz factor in AGN jets, particularly the relativistic jets in Blazars, is by observing the apparent speed of the jet and estimating the Doppler beaming factor. The apparent speed of the jets can be calculated directly by using Very Long Baseline Interferometry (VLBI) observations \citep[eg.][]{j01,k04,b08}, but it is much more complicated to estimate the Doppler beaming factor \citep[see][for a comparison of different methods]{lv99}. They have found that a typical radio-loud quasar has a Lorentz factor $>$ 10, while a typical BL Lac object has a Lorentz factor $>$ 5.

Many studies have tried to constrain the Lorentz factor distribution of the AGN population as a whole, instead of estimating the Lorentz factor for individual sources. A power law form of Lorentz factor distribution has been seen for the most relativistic AGN jets, mainly found in blazars. \cite{pu92} found a Lorentz factor distribution in the form of N($\Gamma$) $\propto \Gamma^{-2.3}$ while \cite{lm97} used a distribution of the form N($\Gamma$) $\propto \Gamma^{\alpha}$ where $-1.5 < \alpha < -1.75$. Recently, \cite{s16} constrained a distribution of N($\Gamma$) $\propto \Gamma^{-2.1\pm0.4}$ in the $\Gamma$ range of 1 to 40 using the optical Fundamental Plane of black hole activity \citep{s15}. It is interesting to note that the parent Lorentz factor distribution for BHXBs obtained from this study is quite similar to the Lorentz factor distributions expected for AGN, and follows the relation N($\Gamma$) $\propto \Gamma^{~-1.88^{+0.27}_{-0.34}}$.

\subsection{X-ray flux and Lorentz factor}

Assuming that the IR excess is caused by the onset of a jet in the hard state, it is also important to check if there is a correlation between the observed IR excess (or indirectly the Lorentz factor) and the X-ray luminosity of the source during the transition. If such a correlation exists, then it is important to normalize the IR excess of these sources by first removing the effect of having different X-ray luminosity during the transition, in order to correctly use the IR excess to constrain the Lorentz factor of these sources. To check this, we use the average X-ray luminosity of our sample in the 2-10 keV energy band, calculated from the X-ray fluxes observed during the start and the end of the infrared transition.

\begin{figure}
\center
\includegraphics[height=0.35\textwidth]{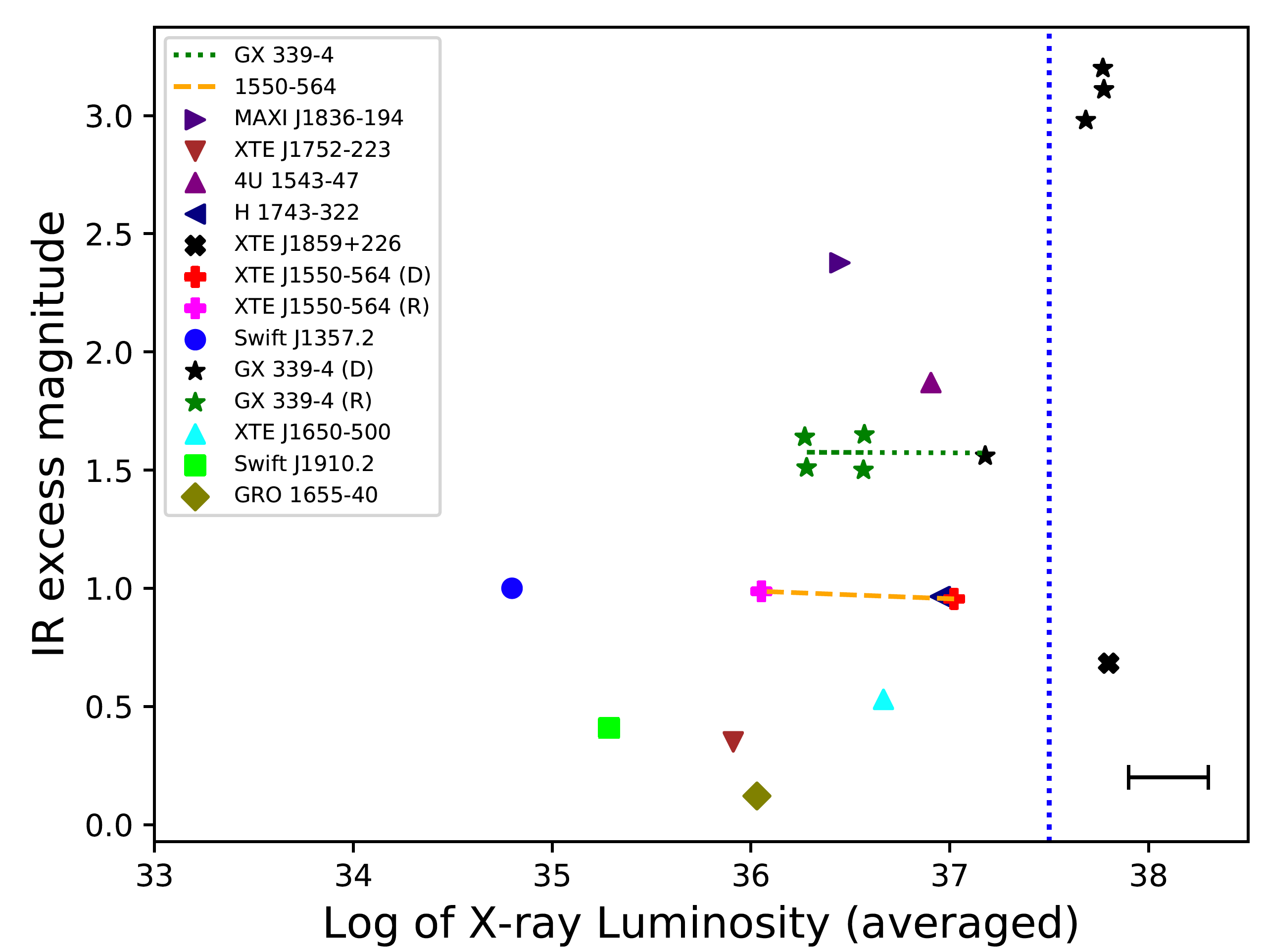}
\caption{IR excess dependence on X-ray luminosity (2-10 keV) for XTE J1550-564, GX 339-4, and all the other BHXB sources in our sample. As seen for XTE J1550-564 (shown as plus signs, red for IR excess seen during rise (R) i.e. if the IR excess is measured during SS to HS transition, and pink during drop (D) i.e. when measured during HS to SS transition), we see that the same IR excess is seen in both the rise and the drop of the source. On the other hand for GX 339-4 (shown here in green stars for rise, and in black stars for drop), there are two ranges of IR excess observed at different X-ray luminosity ranges. The blue dotted line is plotted to illustrate the difference at higher luminosity. A typical error-bar representing the uncertainty of the X-ray luminosity, owing to the variability of X-ray flux during the infrared transition, is shown for reference in the bottom right part of the plot.}
\end{figure}

We first consider the case of GX 339-4 as this source has a fair representation of many IR excess observed during various state transitions over the years. We measured the IR excess observed for GX 339-4 in 8 different state transitions (four during the transition to the soft state and another four during transition to hard state). We see that the IR excess measured when the source is transiting to the hard state is always in a similar range ($\sim$ 1.5 magnitude change in IR). But on the other hand, the IR excess measured during the transition to the soft state varies with respect to the X-ray luminosity of the system. For smaller X-ray luminosities, the IR excess is small ($\sim$ 1.5 mag, similar to the IR excess measured during the transition to the hard state), while for higher X-ray luminosities the IR excess measured is much higher ($\sim$ 3.2 mag). So we see two different ranges of IR excess for GX 339-4, but there is no single trend correlating the X-ray luminosity (2-10 keV) to the IR excess observed (see Fig 9). We find that using a larger X-ray energy range of 0.1--100 keV (for sources with available data) does not significantly change the results.

XTE J1550-564 is the only other source for which we had a measurement of IR excess during both the rise and the drop of the source. It is interesting to note that the IR excess seen in both these transitions are remarkably similar, although the X-ray luminosities during these two transitions were different. So we cannot find a correlation between the X-ray luminosity and the observed IR excess for XTE J1550-564. Although for higher X-ray luminosity ranges we saw that GX 339-4 has a different range of IR excess, all the other BHXBs in our sample do not have such high X-ray luminosity, except XTE J1859+226. For XTE J1859+226, the inclination is $\sim$ 60 degrees. As shown in Fig 1, this inclination is neither too high nor too low to have much effect on IR excess for different Lorentz factors. Hence we find no evidence for a dependency of X-ray luminosity, except above 10$^{\sim37.5}$ erg s$^{-1}$ (2-10 keV). Therefore we do not need to normalize the IR excess for any possible effect of having different X-ray luminosities during the transition. We note that generally at the hard-to-soft transition, the IR emission as well as the hard X-ray flux drop simultaneously, before the drop in the radio wavelengths \cite[see e.g. Fig. 1 in][]{ho05}. On the other hand at the soft-to-hard transition, the radio and the hard X-ray flux rise first, followed by the IR after a significant delay \citep[$\sim$10-12 days in the case of GX339-4, see e.g.][]{c13}. The date range of the IR rise over this transition may not therefore correspond to the dates of the X-ray transition exactly. However, changing this date range would only serve to shift some of the data points slightly in the horizontal direction in Fig. 9, which does not change the conclusions here, and the amplitude of the IR excess remains the same. As there are only two sources in our sample for which we have data on both IR drop and rise, our data is currently insufficient to test for any dependency of the IR drop or rise and the X-ray luminosity at which this occurs, on other physical processes like the internal jet evolution during the transition. With more data, we could test this in a follow-up work.

It is unclear why the IR excess of GX 339--4 appears to have two populations ($\sim 1.5$ mag and $\sim 3$ mag). However if, as our model assumes, relativistic beaming is responsible for the amplitude of the IR excess, it could be suggested that the Lorentz factor increases for GX 339--4 at these very high luminosities of $10^{38}$ erg s$^{-1}$. This would require the inclination to be low (see below) such that the jet is pointing towards us. At these high luminosities, at the point in which the source is making a transition towards the soft state, the inner radius of the disc is moving to smaller radii and the hot flow region is shrinking \citep[e.g.][,and references therein]{femu}. This may result in the Lorentz factor increasing \citep[see][]{fe04,tr2014} and could be caused by the black hole spin playing a role in boosting the jet velocity, although this is speculative. For all other X-ray luminosities (below $10^{37.5}$), we find no evidence for a relation between the Lorentz factor and the luminosity, but we do find evidence (in the only two BHXBs with more than one measurement) of a constant value for the Lorentz factor at different X-ray luminosities. Hence, although both the jet flux and disc flux increase with mass accretion rate, this analysis shows that the Lorentz factor changes the jet flux, giving a boost to some and a reduction to others, compared to the disc flux. The Lorentz factor and the inclination angle appears to define how bright the jet is compared to the disc, at any given X-ray luminosity.

\begin{figure}
\center
\includegraphics[height=0.33\textwidth]{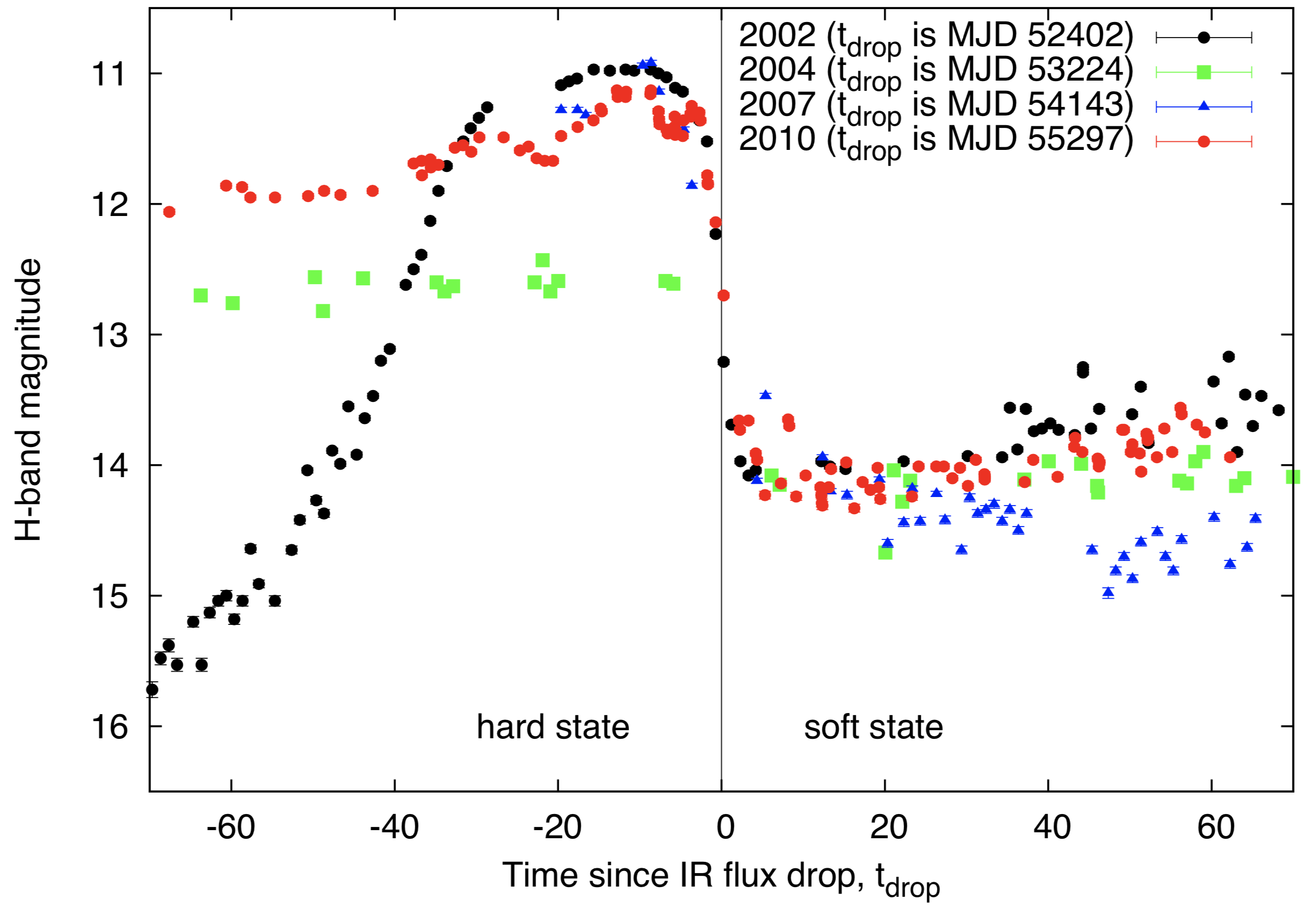}
\includegraphics[height=0.33\textwidth]{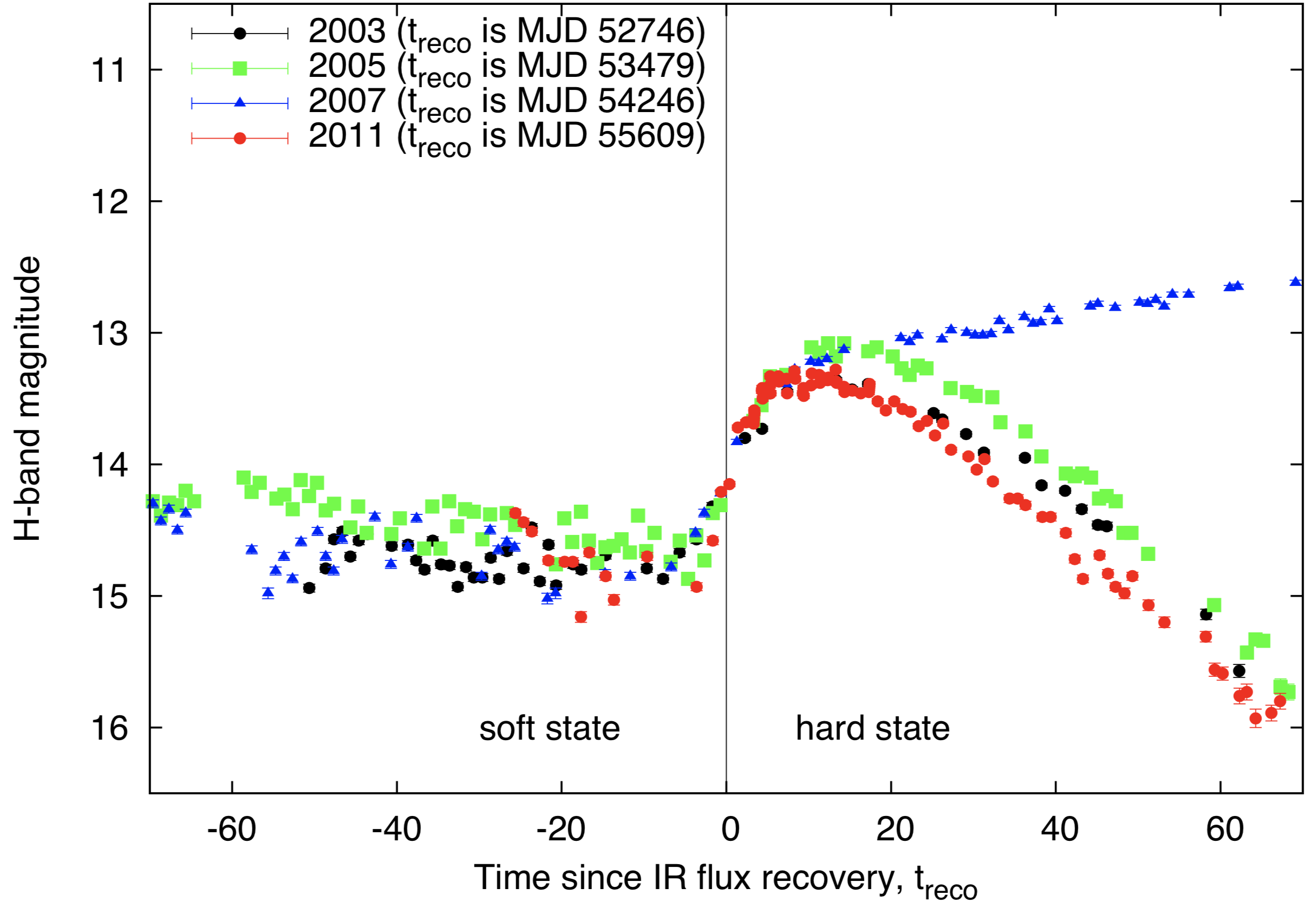}
\caption{Overall light curves of all the times when GX 339-4 went into a transition and showed an IR excess/drop in the spectra.}
\end{figure}

\subsection{IR excess and the inclination of GX 339-4}

GX 339-4 is one of the best studied BHXBs at IR wavelengths. There are four transitions of this source which were properly covered in the IR regime (2004, 2005, 2007 and 2010), during both the rise and the decay phase. A combined plot of all the times when this source went into a transition and showed an IR excess/drop in the spectra is shown in Fig 10. As shown, the H-band magnitude before the transition to the hard state is almost the same for all the four outbursts. Similarly, after the compact jet switches off and the source transits to the soft state, the H-band magnitude drops back to the same value for all four outbursts, irrespective of the H-band magnitude during the hard-state. If we take the behavior of GX339-4 to be typical of BHXBs, then we can say that it is better to calculate the IR magnitude change during the soft to hard transition, as the data seems to be more consistent compared to the flux drop during the hard to soft transition (this may only be an issue at very high X-ray luminosities; see Section 5.4).

The inclination of GX 339-4 is very weakly constrained. \cite{c2002} makes the argument that the orbital inclination has to be less than 60$^\circ$ from the lack of eclipses present in the optical data. \cite{ludlam} argues that the inclination cannot be lower than $\sim$40$^\circ$ in order to have a dynamical mass consistent with the findings of \cite{hygx}. \cite{yamada} found a possible range of inclination 25$^\circ$ -45$^\circ$ at 90\% confidence level. But, \cite{wu2001} carried out optical spectroscopic observations of GX 339-4 during its high-soft and low-hard X-ray spectral states and found that the orbital inclination is about 15$^\circ$ if the orbital period is 14.8 hours. While \cite{kdone} finds that inclinations i $>$ 45$^\circ$ give better fits in the high-soft state when fitting the disk continuum to measure the spin, \cite{s2011} analyses the iron-K line with a diskline model to arrive at a best-fit inclination angle of $\sim$50$^\circ$. Most recently, \cite{h17} analysed the radial velocity curve and projected rotational velocity of GX 339-4 and constrain the binary inclination to  37$^\circ$< i < 78$^\circ$. Our analysis predicts that the inclination of GX 339-4 should be much lower than what were previously expected by many studies, as the observed IR excess of GX 339-4 is consistent with our model only if i $\lesssim 15^\circ$ (see Fig. 3).

\subsection{Caveats of our study and errors involved}

It is important to note the caveats in our study, which could contribute to the uncertainty in the result of this paper, and possibly increase the free parameters in our model. The physical parameters of our sample, including the masses of the compact object and companion star, the inclination angle of the source with respect to the line of sight and the orbital period, etc, are collected from the literature. Different methods (mainly using optical or radio observations) have been used to estimate the inclination of the sources, which is the major source of error. For the inclination obtained from optical studies, we have assumed that the jet is perpendicular to the orbital plane. Indeed, \cite{fragos} predict that the majority of BHXBs have rather small ($<$ 10$^\circ$) misalignment angles. But this might not be the case for all the sources in our sample, as there are BHXBs where the jet definitely appears to be misaligned with the binary orbit \citep[eg.][]{macca02}. Furthermore, the prevalence of Type C QPOs in BHXBs suggests that if the relativistic precession model is correct \citep[see eg.][]{stella}, then the jet should be launched from the inner disk and hence misalignments between the binary orbit and inner disk may be common. 

Another variable that we have not accounted for in our model is the geometry/velocity profile of the jet. If the jet is not strictly conical but rather flared, then adiabatic expansion should kill off the low-frequency emission faster than expected, giving a more inverted spectrum. Similarly, if the jet velocity profile is not constant, but the jet accelerates on moving outwards, it could beam the emission more/less at lower/higher frequencies, again affecting the spectral shape. A more inverted spectrum would imply higher jet emission relative to the disk, and vice versa for a steeper spectrum.

We also use the assumption that the whole projected disc of the BHXB contributes towards IR emission. This might not be completely true because probably only the outer part of the disc will emit in the IR regime. However, since we are calculating the `relative' IR excess for all these sources, we do not expect it to be a big issue in our analysis. Moreover, by comparing the quiescence IR magnitude of the companion star to the faintest IR magnitude of the black hole system during the transition, we find that the contribution from the companion star is much less than 10\% of the total flux, except for the case of GRO 1655-40 (see Section 2.7). Hence we neglect the IR emission coming from the companion star in our model, and subtract its contribution for GRO J1655-40.

Due to lack of enough IR monitoring of BHXBs during state transitions, while some of the IR flux changes have been calculated during the IR rise (i.e. when the source is transiting from soft to hard state), few others were estimated during the IR decay period (i.e. when the source transitions to the soft state and the jet switches off). This could also introduce additional uncertainty in our result. Additionally, we use both H-band and K-band data to measure the IR excess. It is important to note that, amplitude of IR excess measured using K-bands might be slightly greater than the excess measured in H-band. Moreover, adequate IR coverage is not available for all of the sources. So while estimating the IR flux change during state transitions, the data available for some sources were severely limited and could give rise to additional uncertainties. We attempted to include such errors in the uncertainty of the observed IR excess, but better measurements are needed for future studies. 

Finally, the Bayesian modelling involves making a slew of assumptions, the most important being
that the Lorentz factors are drawn from a parent distribution common to BHXBs, and that this
parent distribution has the form of a power law with index $\alpha$ as a single free
(hyper-)parameter. This choice seems reasonable given that the Lorentz factors in AGN jets 
also appear to follow a power law distribution. However, if this assumption is incorrect, then
the results of our modelling should be taken with caution. Furthermore, we have not explored other possibilities 
for the parent distribution.

\section{Conclusion}

The NIR emission seen in BHXBs is expected to originate mainly at the outer part of the accretion disc, and from the jet. An excess of IR emission is observed in many BHXBs in the hard state. We study this excess in IR emission, using a compilation of all the BHXBs in the literature for which an IR excess has been measured. Using a simple qualitative model for beaming of IR jet emission, and for the projected area of the accretion disc, we show that the amplitude of the IR fade or recovery over state transitions is expected to be very low for intermediate inclination angles (~30 - ~60 deg). The observations confirm that within this range of inclination angle, there is no BHXB with a prominent IR excess. Using the amplitude of the IR fade/recovery, the known orbital parameters and a simple Bayesian framework, we constrain for the first time the Lorentz factor of jets in several BHXBs. Under the assumption that the Lorentz factor distribution for BHXB jets is a power-law N($\Gamma)\propto \Gamma^{\alpha}$, we find that $\alpha~=~-1.88^{+0.27}_{-0.34}$, which is remarkably similar to the index of the power-law for the bulk Lorentz factor distributions of highly relativistic jets in AGN. This work can be improved using better inclination measurements and adequate IR monitoring of more BHXBs during state transitions. We also find that the very high amplitude IR fade/recovery seen repeatedly in GX 339-4 requires a much lower inclination angle ($<$ $\sim$ 15$^\circ$) than previously expected by many studies. These results demonstrate how useful OIR monitoring over state transitions is for studying jet properties.\\

\acknowledgments

ACKNOWLEDGEMENT\\

DMR thanks the International Space Science Institute (ISSI) in Bern, Switzerland for support and hospitality for the team meeting `Looking at the disc-jet coupling from different angles: inclination dependence of black-hole accretion observables' in October--November 2018. DMB and DMR acknowledge the support of the NYU Abu Dhabi Research Enhancement Fund under grant RE124.

\bibliography{template}

\end{document}